\documentclass[12pt]{article}
\usepackage{amsmath, amssymb}
\usepackage{amsfonts}
\usepackage{graphicx,psfrag,epsf}
\usepackage{enumerate}
\usepackage{natbib, xcolor}
\usepackage{rotating}
\usepackage{url} 
\usepackage[noend]{algpseudocode}
\usepackage{algorithm}
\usepackage{booktabs}
\usepackage{multirow}
\usepackage{verbatim}
\usepackage{array}
\usepackage{caption}
\usepackage{setspace}
\usepackage{fixltx2e}
\usepackage[compact]{titlesec}

\newcommand{\blind}{0}

\newcommand{\bbeta}{\mbox{\boldmath $\beta$}}
\newcommand{\hatbbeta}{\mbox{$\boldsymbol{\skew{2}\hat\beta}$}}

\newcommand{\bmu}{\mbox{\boldmath $\mu$}}
\newcommand{\bpsi}{\mbox{\boldmath $\psi$}}

\newcommand{\bSigma}{\mbox{\boldmath $\Sigma$}}

\newcolumntype{L}[1]{>{\raggedright\let\newline\\\arraybackslash\hspace{0pt}}m{#1}}
\newcolumntype{C}[1]{>{\centering\let\newline\\\arraybackslash\hspace{0pt}}m{#1}}
\newcolumntype{R}[1]{>{\raggedleft\let\newline\\\arraybackslash\hspace{0pt}}m{#1}}
\algnewcommand\And{\textbf{and}}

\DeclareMathOperator*{\argmin}{arg\,min}
\newtheorem{theorem}{Theorem}

\begin{document}



\def\spacingset#1{\renewcommand{\baselinestretch}%
{#1}\small\normalsize} \spacingset{1}

\if0\blind
{
  \title{\bf Robust Updating of a Risk Prediction Model by Integrating External Ranking Information}
  \author{Nicholas C. Henderson$^{1}$ \\
$^{1}${\small Department of Biostatistics, University of Michigan} } 
\date{}
  \maketitle
} \fi

\if1\blind
{
  \bigskip
  \bigskip
  \bigskip
  \begin{center}
    {\LARGE\bf Robust Updating of a Risk Prediction Model by Integrating External Ranking Information  \\[2mm]  }
\end{center}
  \medskip
} \fi

\bigskip
\begin{abstract}
Utilizing established risk factors and prognostic models
can often improve the construction of a newer risk model
that uses novel biomarkers in a smaller, internal study.
However, directly borrowing information from an established prognostic model 
is often unsuitable due to differences in study populations, patient outcomes measured, and other specific features of the internal study design.
To better enable the use of established prognostic information when constructing
a novel risk model, we propose an estimation approach centered around the 
idea that the risk rankings rather than the risk scores
from an established prognostic model are often more transportable
to the internal study context. To leverage external ranking information, our approach introduces the ranking parameters associated with 
the regression coefficients of an internal risk model and estimates
the internal risk model parameters by penalizing a ranking-based discrepancy measure between the ranking parameters and the rankings implied by the established prognostic model. Our method does not require the external prognostic model to have a specific form, but only requires one to compute risk score rankings from an external model.
Simulation studies demonstrate that our method leads to competitive predictive performance  and performs particularly well when the true internal and external prognostic models have high rank correlation but large discrepancies between their underlying risk scores. We demonstrate the use of our approach through the development of a prognostic model for advanced prostate cancer patients who were treated with an immune checkpoint inhibitor.
\end{abstract}

\noindent%
\vfill

\setlength{\parskip}{.4cm}
\setlength{\parindent}{0in}

\newpage
\spacingset{1.65}

\section{Introduction}

The growing availability of large-scale datasets and statistical summaries from such datasets
has led to increased interest in incorporating information from these sources when conducting
a new study. These much larger datasets include sources such as disease registries,
census data, and published risk models. Such data sources can often be leveraged to 
improve inferences about questions of interest in the new ``internal'' study that has 
a much smaller sample size than these external data sources.
In the data integration literature that focuses on this setting (e.g., \cite{chatterjee2016}), the data 
integration task is often formulated in the following way. A researcher
has information available from a large ``external'' data source that describes the relationship
between an outcome variable $Y$ and an associated set of covariates $\mathbf{Z}$, and the
researcher also collects the same outcome variable $Y$ and covariates $\mathbf{Z}$ in a much 
smaller ``internal'' dataset that the researcher is using
for the study of interest. In addition, the internal study also records information on an additional
collection of covariates $\mathbf{B}$ which is not available from the external data source.
In this context, the main goal of data integration methods is then to harness estimates of features 
of the joint distribution of $(Y, \mathbf{Z})$ from the external data source in order to improve
inferences about the distribution $(Y, \mathbf{Z}, \mathbf{B})$ within the smaller, internal
study of interest.

There has been much recent activity in developing statistical methods to perform data integration
between a large, external data source and a smaller, more focused internal study. To briefly
highlight several prominent and relevant examples from this literature, we note that  
\cite{chatterjee2016} proposed a constrained maximum likelihood framework for 
combining individual-level data from an internal study with summary-level information
from an external data source. Other methods developed for a similar data setting
include \cite{qin2015}, \cite{estes2018}, \cite{cheng2018}, and \cite{gu2019},
and methods which focus on integrating information from multiple, possibly heterogeneous,
external sources include \cite{antonelli2017} and \cite{gu2023}. Other related methods
have focused more on incorporating information from external risk models
where the covariate distributions between the external and internal covariate
distributions are notably different, \citep{han2019, taylor2023},
or where little information exists on how the external risk model utilizes patient covariates \citep{han2023}.

A limitation of many of the methods listed above is that they are often not well-equipped
to handle applications where the external data source has considerable relevance 
to the objectives of the smaller, internal study but key study differences 
prevent direct application of a data integration procedure. As an example of this,
suppose the internal study collects data of the form $(Y, \mathbf{Z}, \mathbf{B})$
while the external model is a risk model characterizing the distribution
of an outcome $\widetilde{Y}$ given $\mathbf{Z}$, and the internal data and 
external model outcomes ($Y$ and $\widetilde{Y}$) are well known to have some form
of positive association though the precise magnitude of the association may not 
be as well known and can depend on particular features of the internal study context.
For example, suppose the purpose of the internal study is to investigate the role
of a newer set of biomarkers in predicting prostate cancer progression and the internal
data is structured so that response in the observed prostate-specific antigen (PSA) \citep{schroder2000} is the 
most sensible measure to use as the primary outcome. To better characterize the role of these novel biomarkers 
in prostate cancer, the internal study investigators may want to incorporate
results from external prostate cancer risk models that have been 
constructed and validated on much larger datasets. However, one drawback of this
is that such risk models may have been built on completely different outcomes
such as progression-free survival (PFS), and hence, in this context, directly calibrating
the internal risk model to the external risk model or shrinking 
parameter estimates from the internal risk model towards parameters of the external 
risk model is likely not a suitable approach. 
Nevertheless, although direct integration of estimates from the external risk model
may not work in this context, risk scores from the external model may still be highly
relevant for explaining variation in internal-data outcomes and incorporating
such external information can improve inferences about the main aims of the internal study.

\begin{figure}
\centering
     \includegraphics[width=5.5in,height=4.0in]{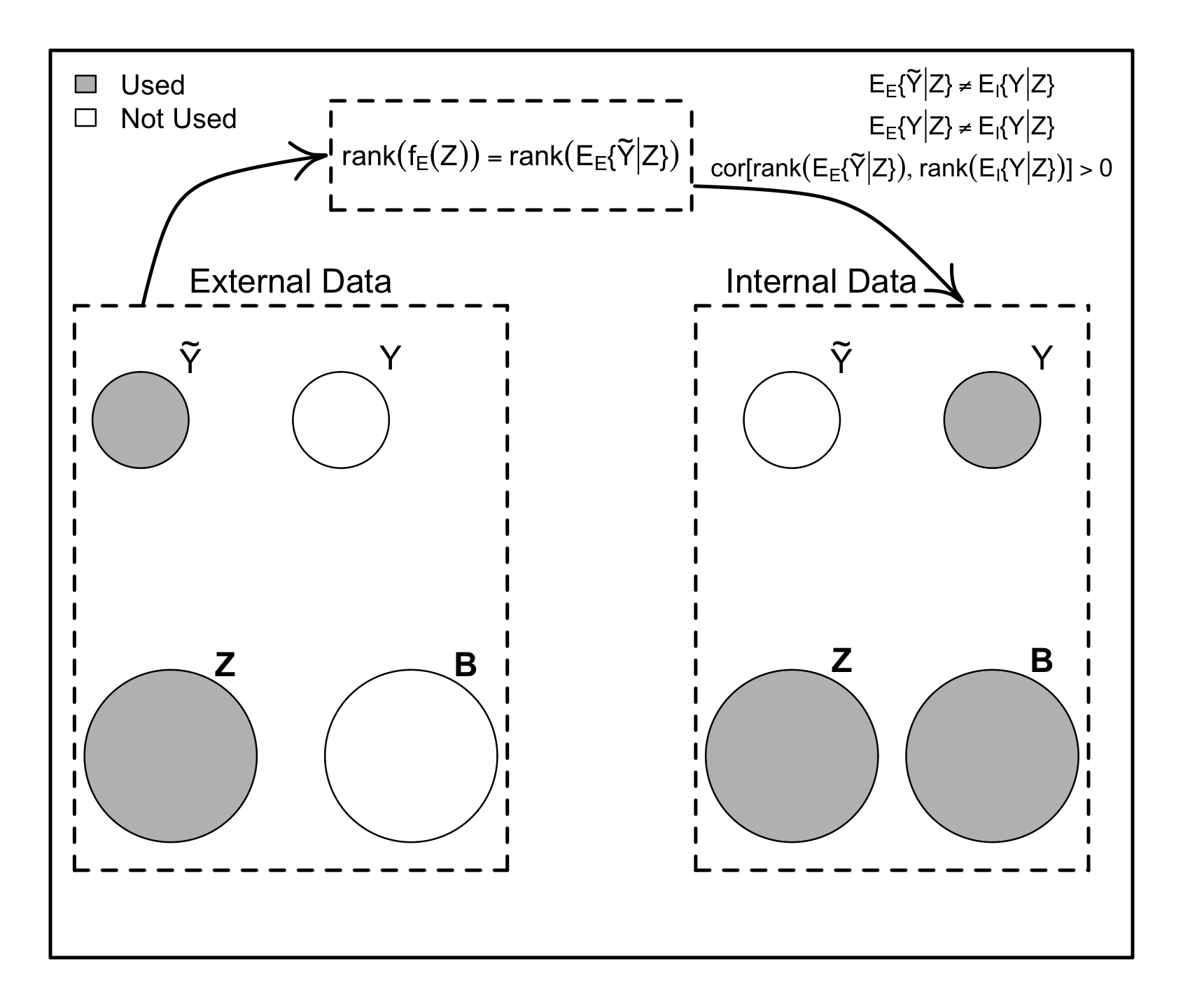}
\caption{Schematic depicting which quantities are captured in the internal and external data sources and our transportability assumptions.}
\label{fig:schematic}
\end{figure}

When key differences between the internal and external data sources are present and 
make direct integration of study estimates inappropriate, the ranking information from the external data source is often better preserved
and will often be a better feature of the external data to utilize
when performing data integration. To be more concrete, suppose the external
data source has information related to the expectation of an outcome
$\widetilde{Y}$ given covariates $\mathbf{Z}$ (denote this
expectation by $f_{E}(\mathbf{Z}) = E_{E}\{ \widetilde{Y} \mid \mathbf{Z} \}$),
while the aim of the internal study is to estimate
the conditional expectation $f_{I}(\mathbf{Z}, \mathbf{B}) = E_{I}\{ Y \mid \mathbf{Z}, \mathbf{B} \}$ for an outcome $Y$ in the internal-study population. 
In this context, 
one can often expect that the rankings of $f_{E}(\mathbf{Z})$ will have a
greater degree of transportability to the internal study setting.
Figure \ref{fig:schematic} has a schematic summarizing a possible scenario
of how the external and internal data sources are related 
and what main assumptions we would be making in this setting.
As shown in Figure \ref{fig:schematic}, the external and internal data
do not use the same outcome, but we will still use the rankings
of the conditional expectations $E_{E}\{ \tilde{Y} \mid \mathbf{Z} \}$ of the outcomes
in the external data context to
inform the internal study.
Letting these rankings inform the internal study is based on
the assumption that, despite the differences in the internal
and external study contexts, 
the internal and external conditional expectations 
$E_{E}\{ \widetilde{Y} \mid \mathbf{Z} \}$ and $E_{I}\{ Y \mid \mathbf{Z} \}$
will have some degree of rank-based positive association.

To incorporate external ranking information into an internal study, 
our proposed approaches focus on estimation with ranking-based penalization.
Adopting a penalization framework allows for borrowing of information
from the external data source without requiring exact calibration
of the rankings to match the rankings implied by the external model. 
Often, the contexts of the internal and external studies differ enough that 
any form of exact calibration may be undesirable, and penalization
provides more flexibility in controlling the level of borrowing
by letting the user specify the magnitude of the penalty hyperparameter.
To target only rankings in the penalization, we propose penalty functions
that only penalize the degree to which the rankings from the internal 
risk model deviate from the rankings that the external risk model 
would assign to the internal-data observations, and the penalty
functions do not directly penalize the magnitude of the internal-model
parameters in any way. Specifically, our ranking penalties most closely
resemble classical measures of rank correlation and rank-based objective
functions used in robust regression approaches. While such rank-based
discrepancies have been used as objective functions for robust regression
(e.g., \cite{Han1987}, \cite{johnson2008}, \cite{sievers2004}, \cite{sun2024}),
their use as penalty functions to perform more robust data integration
from external models has, to our knowledge, received almost no 
attention in the data integration literature.

This paper is organized as follows. Section \ref{sec:model} describes the data structure of interest and introduces the notion of the ranking parameters in an internal risk model. Section \ref{sec:model} also introduces association measures between these ranking parameters and an arbitrary collection of external rankings.
Section \ref{sec:penalized_estimation} details a novel penalized regression procedure for estimating
the regression parameters of an internal risk model where the penalization reflects the degree to which
the internal model ranking parameters deviate from risk score rankings associated with an external risk model.
Section \ref{sec:sim_studies} presents the results of two simulation studies evaluating the ability of our ranking penalized procedure to improve the predictive performance of estimated internal risk models.
In Section \ref{sec:application}, we demonstrate the use of our approach for updating an established risk model using an example involving prostate cancer patients receiving an immune checkpoint inhibitor, and we then conclude with a brief discussion in Section \ref{sec:discussion}.

\section{Internal Risk Models and Ranking Parameters} \label{sec:model}
\subsection{Data Structure}
We assume that we have observed $n$ responses $Y_{i}$ with associated covariate vectors $\mathbf{x}_{i}$ from an internal dataset, which we denote with  
\begin{equation}
\mathcal{D}_{I} = \{ (Y_{i}, \mathbf{x}_{i}); i=1,\ldots, n\},
\nonumber
\end{equation}
where $Y_{i} \in \mathbb{R}$ and $\mathbf{x}_{i} = (x_{i1}, \ldots, x_{ip})^{T} \in \mathbb{R}^{p}$. The pairs $(Y_{i}, \mathbf{x}_{i})$ in $\mathcal{D}_{I}$ represent an i.i.d. sample from a population with joint cumulative distribution function 
$F_{I}(y, \mathbf{x}) = P\{ Y_{i} \leq y, x_{i1} \leq x_{1}, \ldots, x_{ip} \leq x_{p} \}$. The covariate vectors $\mathbf{x}_{i}$ in $\mathcal{D}_{I}$ are assumed to have the form $\mathbf{x}_{i} = (\mathbf{z}_{i}, \mathbf{b}_{i})$, where $\mathbf{z}_{i} \in \mathbb{R}^{q}$ (with $q < p$) is a collection of covariates that can be used to compute risk scores from an established external risk model, and $\mathbf{b}_{i} \in \mathcal{B} \subseteq \mathbb{R}^{p - q}$ is a collection of new covariates to be used in building a new risk model using the internal dataset $\mathcal{D}_{I}$. 
We will refer to the covariates in $\mathbf{z}_{i}$ as the ``conventional covariates'' and the covariates in $\mathbf{b}_{i}$ as the ``novel covariates''. 

Regarding the stated external risk model, our only assumption about this external model is that it can take a vector $\mathbf{z}$ of conventional covariates as input and return an associated risk score $f_{E}(\mathbf{z})$. In the context of applying the external risk model to the internal dataset $\mathcal{D}_{I}$, larger values of $f_{E}(\mathbf{z}_{i})$ should be interpreted as predicting a larger value of the response $Y_{i}$.

\textbf{Transportability Assumptions.} Our main goal is to use the information provided by this external risk model to improve the construction of a new risk model using the internal dataset $\mathcal{D}_{I}$.
To elaborate further, consider the random 
variables $f_{I}(\mathbf{x}_{i})$ and $g_{I}(\mathbf{z}_{i})$ defined as
\begin{equation}
f_{I}(\mathbf{z}_{i}) = E_{I}\{ Y_{i} | \mathbf{z}_{i} \} \qquad 
\textrm{and} \qquad g_{I}(\mathbf{x}_{i}) = E_{I}\{ Y_{i} | \mathbf{x}_{i} \},
\label{eq:internal_exps}
\end{equation}
where the notation $E_{I}$ is used to denote taking the conditional expectation of $Y_{i}$ given $\mathbf{x}_{i}$ using the joint internal-dataset distribution $F_{I}$.

We do not make the assumption that the internal study and external populations are the same or that 
$f_{I}(\mathbf{z}_{i})$ in (\ref{eq:internal_exps}) should match the external risk score $f_{E}(\mathbf{z}_{i})$.
Rather, the main motivation behind our methodology is the expectation that, in many contexts,
the external risk model will have enough relevance to the internal study context to ensure that $f_{I}(\mathbf{z}_{i})$ (and $g_{I}(\mathbf{x}_{i})$) and 
$f_{E}(\mathbf{z}_{i})$ have some degree of nonparametric association. 
The assumption of positive association between $f_{I}(\mathbf{z}_{i})$ and $f_{E}(\mathbf{z}_{i})$ is a quite weak assumption regarding the transportability between any of the external data sources and the internal dataset. 
More concretely, one can think of the established external risk model as being constructed from data of the form $\{ (U_{j}^{E}, \mathbf{z}_{j}^{E}) \}$, where $U_{j}^{E}$ represents either the same type of outcome as $Y_{i}$ or a surrogate outcome for $Y_{i}$, and where $\mathbf{z}_{j}^{E}$ represents the same types of covariates used in the $\mathbf{z}_{i}$ components of the internal dataset.
From this, an implicit assumption driving our methodology is that, for many common measures of rank correlation, the following quantity
\begin{equation}
\rho_{IE} = \textrm{RankCorr}\Big( E_{I}\{Y_{i}|\mathbf{z}_{i}\}, E_{E}\{ U_{j}|\mathbf{z}_{j}^{E}=\mathbf{z}_{i} \} \Big) \nonumber
\end{equation}
will be positive and large enough to enable improved estimation of an internal risk model by borrowing the ranking information from the external risk model. 

\subsection{External Model Rankings and Internal Risk Model Ranking Parameters} 

\textbf{Rankings from the External Model.} Let $\mathbf{z}_{1}, \ldots, \mathbf{z}_{n}$ be the conventional covariates from the internal dataset so that the risk score assigned by the external risk model to individual $i$ in the internal dataset is $f_{E}(\mathbf{z}_{i})$.
Further, let $r_{1}^{E}(\mathbf{z}_{1}), \ldots, r_{n}^{E}(\mathbf{z}_{n})$ denote the ranks of the risk scores obtained from $f_{E}(\mathbf{z}_{1}), \ldots, f_{E}(\mathbf{z}_{n})$. 
The ranks $r_{i}^{E}(\mathbf{z}_{i})$ can be expressed as
\begin{equation}
r_{i}^{E}(\mathbf{z}_{i}) = \sum_{j=1}^{n} I\left\{ f_{E}(\mathbf{z}_{i}) \geq f_{E}(\mathbf{z}_{j}) \right\}. \nonumber 
\end{equation}

\textbf{Ranking parameters for the internal risk model.} Our overall goal is to estimate individual-level risk scores for the internal dataset, and
we will refer to the model generating such risk scores as the ``internal risk model''. 
We assume that the internal risk model is constructed by estimating parameters $\beta_{0}$ and $\bbeta$ in the conditional mean function from a working model that is assumed to have the form
\begin{equation}
E_{W}\{ Y_{i} | \mathbf{x}_{1}, \ldots, \mathbf{x}_{n}, \beta_{0}, \bbeta \}
= H\Big( \beta_{0} + \mathbf{x}_{i}^{T}\bbeta \Big)
\label{eq:working_conditional_mean}
\end{equation}
In (\ref{eq:working_conditional_mean}), $H$ is a monotonically increasing function. Common working models used in practice include $Y_{i} \sim \textrm{Normal}(\beta_{0} + \mathbf{x}_{i}^{T}\bbeta, \sigma^{2})$, in which case $H(x) = x$, or $Y_{i} \sim \textrm{Bernoulli}(1/\{1 + \exp(-\beta_{0} -\mathbf{x}_{i}^{T}\bbeta)\})$, in which case
$H(x) = 1/\{ 1 + \exp(-x)\}$.

When building an internal risk model using (\ref{eq:working_conditional_mean}), we want to consider the risk score rankings implied by a specific value of the parameters $(\beta_{0}, \bbeta)$, and to improve estimation of the internal risk model, we want to incorporate the rank association between these risk score rankings and the external model rankings.
Under an assumed value of $\beta_{0}$ and parameter vector $\bbeta$,
the risk score ranking assigned to individual $i$ can be thought of as a parameter given by
\begin{equation}
\psi_{i}(\bbeta) = \textrm{rank}\Big[ E_{W}\{ Y_{i} | \mathbf{x}_{1}, \ldots, \mathbf{x}_{n},\bbeta \} \Big],
\end{equation}
where $E_{W}$ denotes taking expectation under the working model whose conditional mean function is defined by (\ref{eq:working_conditional_mean}). That is, if one were to assign risk scores using working mean model (\ref{eq:working_conditional_mean}) with parameters $(\beta_{0}, \bbeta)$, then $\psi_{i}(\bbeta)$ represents the risk score ranking of individual $i$ in the internal study.
Because $H$ in (\ref{eq:working_conditional_mean}) is assumed to be strictly increasing, ranking by the values of $\psi_{i}(\bbeta)$ is equivalent to ranking by the values of $\mathbf{x}_{i}^{T}\bbeta$, and
for this reason, we will express the ``ranking parameters'' $\psi_{1}(\bbeta), \ldots, \psi_{n}(\bbeta)$ in terms of the rankings of the values of $\mathbf{x}_{1}^{T}\bbeta, \ldots, \mathbf{x}_{n}^{T}\bbeta$. Specifically, the ranking parameter $\psi_{i}(\bbeta)$ can be expressed as
\begin{eqnarray}
\psi_{i}(\bbeta) 
&=& \sum_{j=1}^{n} I(\mathbf{x}_{i}^{T}\bbeta \geq \mathbf{x}_{j}^{T}\bbeta) 
= \sum_{j=1}^{n} I\left\{ (\mathbf{x}_{i} - \mathbf{x}_{j})^{T}\bbeta \geq 0 \right\}. 
\label{eq:rankingpar_def}
\end{eqnarray}

For computational reasons, we want to work with a ``smoothed'' version of the ranking parameters $\psi_{i}(\bbeta)$ when conducting estimation of the internal risk model. To this end, we define smoothed ranking parameters $\psi_{1}^{\nu}(\bbeta), \ldots, \psi_{n}^{\nu}(\bbeta)$ as
\begin{equation}
\psi_{i, \nu}(\bbeta) 
= \sum_{j=1}^{n} g_{\nu}\Big( (\mathbf{x}_{i} - \mathbf{x}_{j})^{T}\bbeta  \Big), 
\label{eq:ranking_pars}
\end{equation}
where $\nu > 0$ is a tuning parameter, and for any $\nu > 0$, $g_{\nu}(\cdot)$ is a cumulative distribution function. Examples of $g_{\nu}(\cdot)$ include  $g_{\nu}(x) = 1/\{ 1 + \exp(-x/\nu) \}$ and $g_{\nu}(x) = \Phi(x/\nu)$. In our implementation, we set $g_{\nu}(\cdot)$ to $g_{\nu}(x) = 1/\{1 + \exp(-x/\nu)\}$. The tuning parameter $\nu$ should be chosen to be relatively small so that $g_{\nu}(x)$ approximates the indicator function, and strategies for choosing $\nu$ are discussed in Section \ref{ss:computation}.

\textbf{Marginalized ranking parameters.} 
Because we can roughly think of the external risk model ranks $r_{1}^{E}(\mathbf{z}_{1}), \ldots, r_{n}^{E}(\mathbf{z}_{n})$ as providing estimates of the expected rankings of the internal-dataset outcomes $Y_{i}$ given only the conventional covariates $\mathbf{z}_{i}$, ranking parameters that can be interpreted as representing a reported ranking conditional only on the conventional covariates have a stronger justification.
To this end, we define the following alternative ranking parameters 
$\tilde{\psi}_{1}(\bbeta), \ldots, \tilde{\psi}_{n}(\bbeta)$
\begin{equation}
\tilde{\psi}_{i}(\bbeta) = E\Bigg( \textrm{rank}\Big[ E_{W}\{Y_{i}|\mathbf{x}_{1}, \ldots, \mathbf{x}_{n}, \beta_{0}, \bbeta \} \Big| \mathbf{z}_{1}, \ldots, \mathbf{z}_{n}, \bbeta \Big] \Bigg), 
\label{eq:marg_parameters}
\end{equation}
which we refer to as the ``marginalized" ranking parameters.
Letting  $\bbeta = (\bbeta_{z}^{T}, \bbeta_{b}^{T})^{T}$, the relationship between $\tilde{\psi}_{i}(\bbeta)$ and $\psi_{i}(\bbeta)$ is given by
\begin{align}
\tilde{\psi}_{i}(\bbeta)
&= E\{ \psi_{i}(\bbeta) \mid \mathbf{z}_{1}, \ldots, \mathbf{z}_{n}, \bbeta\} \nonumber \\
&= E\Big[ \sum_{j=1}^{n}I\{ (\mathbf{z}_{i} - \mathbf{z}_{j})^{T}\bbeta_{z} + (\mathbf{b}_{i} - \mathbf{b}_{j})^{T}\bbeta_{b} \geq 0 \} \Big| \mathbf{z}_{1}, \ldots, \mathbf{z}_{n}, \bbeta \Big].
\label{eq:margranking_full}
\end{align}

It is worth noting that the set of marginalized ranking parameters 
$\tilde{\psi}_{1}(\bbeta), \ldots, \tilde{\psi}_{n}(\bbeta)$ may not constitute a collection of ranks themselves in the sense that they will not be a permutation of the set of integers $\{1,\ldots, n\}$. Nevertheless, our main aim is to utilize some degree of appropriate nonparametric association between the internal model risk scores $\mathbf{x}_{i}^{T}\bbeta$ and the external model risk rankings, and using (\ref{eq:marg_parameters}) does effectively capture this association in practice, even though the $\tilde{\psi}_{i}(\bbeta)$ are not strictly a collection of ranks.

In practice, we use smooth versions of the marginalized ranking 
parameters (\ref{eq:margranking_full}) by replacing the indicator function with a smooth function $g_{\nu}$ and by approximating the expectation in (\ref{eq:margranking_full}) by sampling $\mathbf{b}_{i}$ from an estimated conditional 
distribution of the novel covariates given the conventional covariates.
Specifically, if we have $S$ draws from an estimated conditional distribution
of $\mathbf{b}_{i}$ given $\mathbf{z}_{i}$, we define the smoothed marginalized ranking parameters $\tilde{\psi}_{i}^{\nu}(\bbeta)$ as
\begin{equation}
\tilde{\psi}_{i}^{\nu}( \bbeta )
= \sum_{j=1}^{n} \frac{1}{S} \sum_{s=1}^{S} g_\nu\Big( (\mathbf{z}_{i} - \mathbf{z}_{j})^{T}\bbeta_{z} + (\mathbf{b}_{i}^{(s)} - \mathbf{b}_{j}^{(s)} )^{T}\bbeta_{b} \Big), 
\label{eq:marg_ranking_pars_sample}
\end{equation}
where $\mathbf{b}_{i}^{(s)}$ denotes the $s^{th}$ draw of $\mathbf{b}_{i}$ from an assumed
conditional distribution of $\mathbf{b}_{1}, \ldots, \mathbf{b}_{n}$ given $\mathbf{z}_{1}, \ldots, \mathbf{z}_{n}$.


\subsection{Association Measures between the Internal and External Risk Rankings.}
\label{ss:assoc_measures}

When estimating an internal risk model, we would like our estimation procedure to incorporate comparisons between the internal risk score ranking parameters with the rankings from the external risk model. To conduct such comparisons, we will utilize several measures of association between collections of rankings.

\textbf{Association based on Spearman rank correlation.} 
A classic measure of association between two vectors of ranks is Spearman's rank correlation. Spearman's rank correlation $\rho\left( \bpsi(\bbeta), \mathbf{r}^{E} \right)$ between the vector of ranking parameters $\bpsi(\bbeta) = \left(\psi_{1}(\bbeta), \ldots, \psi_{n}(\bbeta) \right)^{T}$ and the vector of external-model ranks $\mathbf{r}^{E} = \big(r_{1}^{E}(\mathbf{z}_{1}), \ldots, r_{n}^{E}(\mathbf{z}_{n}) \big)^{T}$ is given by
\begin{equation}
\rho\left( \bpsi(\bbeta), \mathbf{r}^{E} \right)
= \frac{ \sum_{i=1}^{n} \psi_{i}(\bbeta)(r_{i}^{E}(\mathbf{z}_{i}) - \bar{r}) }{\sqrt{\sum_{i=1}^{n} \{ \psi_{i}(\bbeta) - \bar{\psi}(\bbeta)\}^{2}}\sqrt{\sum_{i=1}^{n} \{r_{i}^{E}(\mathbf{z}_{i}) - \bar{r}\}^{2}}}, \nonumber 
\end{equation}
where $\bar{\psi}(\bbeta) = \frac{1}{n}\sum_{i=1}^{n} \psi_{i}(\bbeta)$ and $\bar{r} = \frac{1}{n}\sum_{i=1}^{n} r_{i}^{E}(\mathbf{z}_{i})$. If there are no ties among the ranking parameters $\psi_{1}(\bbeta), \ldots, \psi_{n}(\bbeta)$, one can express (see Appendix A) Spearman's rank correlation as
\begin{equation}
\rho\left( \bpsi(\bbeta), \mathbf{r}^{E} \right)
=  
\frac{ 4n^{2} }{\sqrt{(n^{3} - n)/12}\sqrt{\sum_{i=1}^{n} (r_{i}^{E}(\mathbf{z}_{i}) - \bar{r})^{2}}}
\Big[ \frac{1}{4n^{2}}\sum_{i=1}^{n} \psi_{i}(\bbeta)r_{i}^{E}(\mathbf{z}_{i}) - \frac{\bar{r}(n+1)}{8n} \Big]. \nonumber 
\end{equation}

Note that $\rho( \bpsi(\bbeta), \mathbf{r}^{E} )$ only depends on $\bbeta$ through $\sum_{i=1}^{n} \psi_{i}(\bbeta)r_{i}^{E}(\mathbf{z}_{i})$ and that
larger values of $\tfrac{1}{4n^{2}}\sum_{i=1}^{n} \psi_{i}(\bbeta)r_{i}^{E}(\mathbf{z}_{i})$ imply larger values of the Spearman rank correlation. Hence, one measure of the ``concordance'' between $\bpsi(\bbeta)$ and $\mathbf{r}^{E}$
is the sum $\frac{1}{4n^{2}}\sum_{i=1}^{n} \psi_{i}(\bbeta)r_{i}^{E}(\mathbf{z}_{i})$. Using this as motivation, we define the following smooth measure of rank concordance between the smooth ranking parameters and the external risk model rankings
\begin{equation}
D_{Sp,0}^{\nu}(\bbeta; \mathbf{r}^{E}) = \frac{1}{4n^{2}}\sum_{i=1}^{n} \psi_{i}^{\nu}(\bbeta)r_{i}^{E}(\mathbf{z}_{i})
= \frac{1}{4n^{2}}\sum_{i=1}^{n} \sum_{j=1}^{n} r_{i}^{E}(\mathbf{z}_{i}) g_{\nu}\Big( (\mathbf{x}_{i} - \mathbf{x}_{j})^{T}\bbeta \Big).
\label{eq:spearman_discordance}
\end{equation}

\textbf{Association based on Kendall's $\tau$.}
Kendall's tau between the vector of ranking parameters $\psi(\bbeta)$ and the vector of external model rankings $\mathbf{r}^{E}$ is defined as 
\begin{eqnarray}
\kappa( \psi(\bbeta), \mathbf{r}^{E} ) = \frac{1}{n(n-1)} \sum_{i=1}^{n}\sum_{j=1}^{n}  I\Big( \{ \psi_{i}(\bbeta) - \psi_{j}(\bbeta)\}\{r_{i}^{E}(\mathbf{z}_{i}) - r_{j}^{E}(\mathbf{z}_{j}) \} > 0 \Big). \nonumber
\end{eqnarray}
Under an assumption of no ties among the ranking parameters $\psi_{1}(\bbeta), \ldots, \psi_{n}(\bbeta)$, the above expression can be rewritten (see Appendix A) as
\begin{equation}
\kappa( \psi(\bbeta), \mathbf{r}^{E} ) =  C + \frac{2}{n(n-1)}\sum_{i=1}^{n}\sum_{j=1}^{n}   I\Big( (\mathbf{x}_{i} - \mathbf{x}_{j})^{T}\bbeta > 0\Big)\Big\{ I\Big(r_{i}^{E}(\mathbf{z}_{i}) > r_{j}^{E}(\mathbf{z}_{j}) \Big) - 1/2\Big\}, 
\label{eq:kendall_tau_formula}
\end{equation}
where $C$ is a constant that does not depend on $\bbeta$, and using (\ref{eq:kendall_tau_formula}) as motivation, we define the following smooth Kendall's $\tau$-inspired measure of rank concordance 
\begin{equation}
D_{Ke, 0}^{\nu}(\bbeta; \mathbf{r}^{E} ) = \frac{2}{n(n-1)}\sum_{i=1}^{n}\sum_{j=1}^{n}  g_{\nu}\Big( (\mathbf{x}_{i} - \mathbf{x}_{j})^{T}\bbeta \Big)\Big\{I\Big( r_{i}^{E}(\mathbf{z}_{i}) > r_{j}^{E}(\mathbf{z}_{j}) \Big) - 1/2 \Big\}. 
\label{eq:kendall_discordance}
\end{equation}

\textbf{Association measures with marginalized ranking parameters.}
When using the marginalized ranking parameters, we define a Spearman-type concordance measure by replacing $\mathbf{x}_{i}$ in (\ref{eq:spearman_discordance}) with draws $\mathbf{x}_{i}^{(s)}$ from the conditional distribution of $\mathbf{x}_{i}$ given $\mathbf{z}_{i}$. 
Specifically, our Spearman-type concordance measure for marginalized ranking parameters is defined as 
\begin{equation}
D_{Sp, 1}^{\nu}(\bbeta; \mathbf{r}^{E})
= -\frac{1}{4Sn^{2}}\sum_{i=1}^{n} \sum_{j=1}^{n} \sum_{s=1}^{S} r_{i}^{E}(\mathbf{z}_{i})g_{\nu}\Big( (\mathbf{x}_{i}^{(s)} - \mathbf{x}_{j}^{(s)})^{T}\bbeta \Big),
\nonumber 
\end{equation}
where $\mathbf{x}_{i}^{(s)} = (\mathbf{z}_{i}^{T}, (\mathbf{b}_{i}^{(s)})^{T} )^{T}$ and $\mathbf{b}_{i}^{(s)}$ is a draw from the conditional distribution of $\mathbf{b}_{i}$ given $\mathbf{z}_{i}$.
Similarly, one can define a Kendall's $\tau$-type measure of association for marginalized ranking parameters $D_{Ke, 1}^{\nu}$ by replacing $\mathbf{x}_{i}$ with $\mathbf{x}_{i}^{(s)}$ in (\ref{eq:kendall_discordance}).

To summarize, we consider four different possible association
measures $D_{Sp,0}^{\nu}$, $D_{Ke,0}^{\nu}$, $D_{Sp, 1}^{\nu}$, $D_{Ke, 1}^{\nu}$ whose choice
depends on: (a) the choice of association measure (Spearman versus Kendall's $\tau$), 
(b) the choice of ranking parameter type (marginalized versus not marginalized). Henceforth, we will use the
generic notation $D_{\bullet}^{\nu}(\cdot, \cdot)$ to denote that
$D_{\bullet}^{\nu}$ is one of the four possible association measures discussed above.

It is worthwhile to note that each possible choice of $D_{\bullet}^{\nu}$
satisfies $0 < D_{\bullet}^{\nu} < 1$ and, when using the non-marginalized ranking parameters, can be expressed as
\begin{equation}
D_{\bullet}^{\nu}(\bbeta; \mathbf{r}^{E}) =  
\sum_{i=1}^{n}\sum_{j=1}^{n} w_{ij} g_{\nu}\Big( (\mathbf{x}_{i} - \mathbf{x}_{j})^{T}\bbeta \Big),
\label{eq:general_disc_form} 
\end{equation}
where $w_{ij}$ can be thought of as fixed weights whose form depends on the chosen association measure. Specifically, 
$w_{ij} = r_{i}^{E}(\mathbf{z}_{i})/4n^{2}$ when considering the Spearman association measure $D_{Sp, 0}^{\nu}$, and setting $w_{ij} = [2I\big(r_{i}^{E}(\mathbf{z}_{i}) > r_{j}^{E}(\mathbf{z}_{j}) \big) - 1]/[n(n-1)]$ yields the Kendall's $\tau$ association measure $D_{Ke, 0}^{\nu}$.
Similar to (\ref{eq:general_disc_form}), when using marginalized ranking parameters, each choice of $D_{\bullet}^{\nu}$ can be expressed as
\begin{equation}
D_{\bullet}^{\nu}(\bbeta; \mathbf{r}^{E}) =  
\sum_{i=1}^{n}\sum_{j=1}^{n} \sum_{s=1}^{S} w_{ijs} g_{\nu}\Big( (\mathbf{x}_{i}^{(s)} - \mathbf{x}_{j}^{(s)})^{T}\bbeta \Big),
\label{eq:general_disc_form_marg} 
\end{equation}
where $w_{ijs} = w_{ij}/S$.

\section{Estimating the Internal Risk Model} \label{sec:penalized_estimation}
\subsection{Penalized Estimation}
To estimate the regression coefficients of an internal risk model, we propose minimizing a penalized objective function $\ell_{\lambda, \alpha}$ that has the form
\begin{equation}
\ell_{\lambda, \alpha}(\beta_{0}, \bbeta) = L_{I}(\beta_{0}, \bbeta; \alpha) - \lambda \log D_{\bullet}^{\nu}(\bbeta, \mathbf{r}^{E}), 
\label{eq:penalized_obj_form}
\end{equation}
where $\lambda \geq 0$ is a nonnegative tuning parameter and where $L_{I}(\beta_{0}, \bbeta; \alpha)$ is a ``local objective function'' that depends only on $(\beta_{0}, \bbeta)$, a hyperparameter $\alpha \geq 0$,
and observations from the internal dataset. Also, $D_{\bullet}^{\nu}$ is one of the rank-based concordance measures between the internal model ranking parameters and the external ranks described in Section \ref{ss:assoc_measures}. 

In this paper, we describe our approach when the local objective function $L_{I}(\beta_{0}, \bbeta;\alpha)$ is a penalized, negative log-likelihood associated with the assumption that the internal-dataset outcomes arise from a generalized linear model (GLM). 
Specifically, we assume the density function of $Y_{i}$ in the internal data can be expressed as
\begin{equation}
f(Y_{i}; \beta_{0}, \bbeta) = \exp\Big\{ \frac{Y_{i}\theta(\beta_{0} + \mathbf{x}_{i}^{T}\bbeta) - b\{ \theta(\beta_{0} + \mathbf{x}_{i}^{T}\bbeta)\}}{a_{i}(\phi)} + c(Y_{i}, \phi) \Big\}, \nonumber
\end{equation}
where $\phi$ is a dispersion parameter and the function $\theta(\cdot)$ is determined by the mean $E(Y_{i}|\mathbf{x}_{i}, \beta_{0}\bbeta)$ and choice of link function $h$, i.e., $h^{-1}\{ E(Y_{i}|\mathbf{x}_{i}, \bbeta) \} = b'\{ \theta(\mathbf{x}_{i}^{T}\bbeta) \}$.
Under this setup, the penalized objective function (\ref{eq:penalized_obj_form}) for a fixed value of $\phi$ becomes
\begin{eqnarray}
\ell_{\lambda, \alpha}(\beta_{0}, \bbeta) &=& 
\sum_{i=1}^{n} \frac{ b\{ \theta(\beta_{0} + \mathbf{x}_{i}^{T}\bbeta) \}}{ a_{i}(\phi) } 
- \sum_{i=1}^{n} \frac{ Y_{i}\theta(\beta_{0} + \mathbf{x}_{i}^{T}\bbeta)}{a_{i}(\phi)}
+ \frac{\alpha}{2}\sum_{j=1}^{p} \beta_{j}^{2} - \lambda \log D_{\bullet}^{\nu}(\bbeta, \mathbf{r}). \nonumber
\end{eqnarray}

We refer to the vector of regression coefficients $\hatbbeta_{\lambda, \alpha}$ that minimizes the penalized objective function (\ref{eq:penalized_obj_form}) as the Rank-ASociated PEnalized Regression (RASPER) estimate of the regression coefficients.

\subsection{Computation and Choice of $\nu$} \label{ss:computation}

One computational challenge raised by the goal of minimizing (\ref{eq:penalized_obj_form}) is that this penalized objective function will not be convex for many typical choices of the local objective function $L_{I}(\beta_{0}, \bbeta; \alpha)$. To address this, we have developed a general inequality for the penalized objective function $\ell_{\lambda, \alpha}(\beta_{0}, \bbeta)$ that can be used to generate direct majorize-minimize (MM) algorithms for common choices of the local objective function. MM algorithms are numerically stable procedures that we have found to work well in practice. An important feature of an MM algorithm is that it is guaranteed to produce a final estimate of $\bbeta$ that improves the value of $\ell_{\lambda, \alpha}(\beta_{0}, \bbeta)$ relative to the initial value of $\bbeta$ chosen. This implies that, while an MM algorithm does not ensure that one finds a global minimizer of (\ref{eq:penalized_obj_form}), it is guaranteed to always produce an estimate $\bbeta^{*}$ that has a smaller value of the objective compared to the minimizer $\hatbbeta_{\textrm{loc}, \lambda, \alpha}$
of the local objective function, provided that one sets the initial value of the MM algorithm to $\hatbbeta_{\textrm{loc}, \lambda, \alpha}$.

To better describe our computational strategy, we introduce additional notation for a few terms. First, we define the $n^{2} \times p$ matrix $\mathbf{A}_{\nu}$ as $\mathbf{A}_{\nu} = \frac{1}{\nu}\big(  \mathbf{X} \otimes \mathbf{1}_{n} - \mathbf{1}_{n} \otimes \mathbf{X}\big)$, where $\otimes$ denotes the Kronecker product. Similarly, we define the $n^{2} \times 1$ vector $\mathbf{w}$ to be $\mathbf{w} = (w_{11}, w_{12}, w_{13}, \ldots, w_{nn})^{T}$. Using $\mathbf{A}_{\nu}$ and $\mathbf{w}$, we can more compactly express the family of concordance measures defined in (\ref{eq:general_disc_form}) as
\begin{equation}
D_{\bullet}^{\nu}(\bbeta, \mathbf{r}^{E}) =  
\mathbf{w}^{T} g_{1}\big( \mathbf{A}_{\nu}\bbeta \big),
\label{eq:general_vector_form} 
\end{equation}

An MM algorithm for minimizing (\ref{eq:penalized_obj_form}) can often be developed by utilizing the results detailed in Theorem \ref{thm:mm_algorithm} below.
\begin{theorem}
When $\ell_{\lambda, \alpha}(\beta_{0}, \bbeta)$ has the form (\ref{eq:penalized_obj_form}) and $D_{\bullet}^{\nu}(\cdot, \cdot)$ has the form (\ref{eq:general_disc_form}), then for any $\bbeta \in \mathbb{R}^{p}$ and $\bbeta^{(t)} \in \mathbb{R}^{p}$,
there is a constant $C(\bbeta^{(t)})$ not depending on $\bbeta$ such that the following inequality and equality hold 
\begin{eqnarray}
\ell_{\lambda, \alpha}(\beta_{0}, \bbeta) &\leq& L_{I}(\beta_{0}, \bbeta; \alpha) -\frac{\lambda}{2}\bbeta^{T}\mathbf{A}_{\nu}^{T}\mathbf{W}_{t}\mathbf{1} - \frac{\lambda}{2}\bbeta^{T}\mathbf{A}_{\nu}^{T}\mathbf{V}_{t}\mathbf{W}_{t}\mathbf{A}_{\nu}\bbeta + C(\bbeta^{(t)}) \label{eq:mm_upperbd} \\
\ell_{\lambda, \alpha}(\beta_{0}, \bbeta) &=& L_{I}(\beta_{0}, \bbeta^{(t)}; \alpha) -\frac{\lambda}{2}(\bbeta^{(t)})^{T}\mathbf{A}_{\nu}^{T}\mathbf{W}_{t}\mathbf{1} - \frac{\lambda}{2}(\bbeta^{(t)})^{T}\mathbf{A}_{\nu}^{T}\mathbf{V}_{t}\mathbf{W}_{t}\mathbf{A}_{\nu}\bbeta^{(t)} + C(\bbeta^{(t)}), \nonumber
\end{eqnarray}
where $\mathbf{W}_{t}$ and $\mathbf{V}_{t}$ are $n^{2} \times n^{2}$ diagonal matrices. The $k^{th}$ diagonal 
elements of $\mathbf{W}_{t}$ and $\mathbf{V}_{t}$ are  $[\mathbf{w}]_{k}g_{1}([\mathbf{A}_{\nu}\bbeta^{(t)}]_{k})/\{ \mathbf{w}^{T}g_{1}(\mathbf{A}_{\nu}\bbeta^{(t)}) \}$ and $0.5\tanh(-    [\mathbf{A}_{\nu}\bbeta^{(t)}]_{k}/2)/2[\mathbf{A}_{\nu}\bbeta^{(t)}]_{k}$, respectively, where $[\mathbf{w}]_{k}$ and $[\mathbf{A}_{\nu}\bbeta]_{k}$
refer to the $k^{th}$ elements of $\mathbf{w}$ and $\mathbf{A}_{\nu}\bbeta$, respectively.
\label{thm:mm_algorithm}
\end{theorem}

A valid MM algorithm for generating a sequence of iterates $\bbeta^{(0}, \bbeta^{(1)}, \bbeta^{(2)}, \ldots$ guaranteed to reduce the objective function $\ell_{\lambda, \alpha}(\beta_{0}, \bbeta)$ in every iteration is obtained by minimizing the upper bound for $\ell_{\lambda, \alpha}(\beta_{0}, \bbeta)$ in (\ref{eq:mm_upperbd}). When the chosen local objective function is a penalized residual sum of squares, i.e., $L_{I}(\beta_{0}, \bbeta;\alpha) = \frac{1}{2}||\mathbf{Y} - \beta_{0}\mathbf{1} - \mathbf{X}\bbeta||^{2} + \frac{\alpha}{2}||\bbeta||^{2}$,
this MM algorithm leads to a direct, iteratively reweighted least squares procedure, where the update of the regression coefficients in the $t^{th}$ iteration of the procedure takes the form
\begin{equation}
\bbeta^{(t+1)} = \Big( \mathbf{X}^{T}\mathbf{X} + \alpha\mathbf{I}_{p} - \lambda \mathbf{A}_{\nu}^{T}\mathbf{V}_{t}\mathbf{W}_{t}\mathbf{A}_{\nu} \Big)^{-1}\Big( \frac{\lambda}{2}\mathbf{A}_{\nu}^{T}\mathbf{W}_{t}\mathbf{1} + \mathbf{X}^{T}(\mathbf{Y} - \beta_{0}\mathbf{1}) \Big).
\label{eq:mm_update}
\end{equation}

\textbf{Choice of $\nu$.}
When setting a default value of $\nu$, we assume that the columns of the design matrix $\mathbf{X}$ have been standardized to have mean zero and unit variance; that is, $\sum_{i = 1}^{n} x_{ik} = 0$ and $\sum_{i=1}^{n} x_{ik}^{2} = (n - 1)$, for each $k$.
When the covariates are standardized and assumed to be independent, we have $E\{ (\mathbf{x}_{i} - \mathbf{x}_{j})^{T}\bbeta \} = \mathbf{0}$ and $\textrm{Var}\{  (\mathbf{x}_{i} - \mathbf{x}_{j})^{T}\bbeta \} = || \bbeta ||^{2}$, for any $\bbeta$.

For the default choice of $\nu$, we set $\nu$ so that $g_{\nu}(-0.25 ||\bbeta || ) \leq 0.01$ and $g_{\nu}(0.25 ||\bbeta|| ) \geq 0.99$, and the approximate smallest value of $\nu$ that guarantees these inequalities is $\nu = 0.1 ||\bbeta||$.
Because $\bbeta$ is not known in practice, we set $\nu$ to $\nu = 0.1||\hatbbeta_{MLE}||$, where $\hatbbeta_{MLE}$ is the estimate of $\bbeta$ obtained by maximizing the unpenalized local objective function $L_{I}(\beta_{0}, \bbeta; 0)$. The motivation for this choice of $\nu$ is to ensure that $g_{\nu}\big( x \big)$ is a close approximation of the indicator function $I\big((\mathbf{x}_{i} - \mathbf{x}_{j})^{T}\bbeta \geq 0 \big)$ for most values of $(\mathbf{x}_{i} - \mathbf{x}_{j})^{T}\bbeta$ that one would consider when estimating the internal risk model regression parameters. Our choice of $\nu$ implies that, for most plausible choices of $\bbeta$, we will have $g_{\nu}(u_{ij}) \leq 0.01$ for values of $u_{ij} = (\mathbf{x}_{i} - \mathbf{x}_{j})^{T}\bbeta$ more than one-fourth standard deviation less than the mean of $u_{ij}$ while $g_{\nu}(u_{ij}) \geq 0.99$ for values of $u_{ij}$ greater than one-fourth standard deviation from the mean of $u_{ij}$.
Our approach for setting $\nu$ also resembles other approaches that employ smooth logistic approximations of indicator functions, for example, in the context of maximizing a smooth approximation of the area under the ROC curve \citep{ma2005} or 
in the context of optimizing individualized treatment rules (\cite{wu2021, varadhan2016}).

\subsection{Hyperparameter Selection}
\textbf{Leave-one-out cross-validation.} A direct way to select
the hyperparameters $(\lambda, \alpha)$ is through leave-one-out cross-validation. The leave-one-out cross-validation score $\textrm{LOO}(\lambda, \alpha)$ we use is obtained by evaluating the local objective function $L_{I}(\beta_{0},\bbeta;\alpha)$ on each holdout observation while ignoring the contribution of the $L_{2}$ penalty to the local objective function. Specifically, we define $\textrm{LOO}(\lambda, \alpha)$ as
\begin{equation}
\textrm{LOO}(\lambda, \alpha) = \frac{1}{n}\sum_{i=1}^{n} L_{I}^{-i}(\hat{\beta}_{0, \lambda, \alpha}^{-i}, \hatbbeta_{\lambda, \alpha}^{-i}; 0), \nonumber 
\end{equation}
where $L_{I}^{i}$ is the local objective function obtained by only using
the single observation $(y_{i}, \mathbf{x}_{i})$ and $\hatbbeta_{\lambda, \alpha}^{-i} = (\hat{\beta}_{\lambda, \alpha, 1}^{-i}, \hat{\beta}_{\lambda, \alpha, 2}^{-i}, \ldots, \hat{\beta}_{\lambda, \alpha, p}^{-i})^{T}$ is the estimate of the vector of regression coefficients obtained when $(y_{i}, \mathbf{x}_{i})$ has been deleted from the internal dataset.

In our implementation, we evaluate $\textrm{LOO}(\lambda, \alpha)$ on a two dimensional grid 
of points $\{ (\lambda_{j}, \alpha_{k}): j=0, \ldots, J + 1; k = 0, \ldots, K + 1\}$, where both
the $\lambda_{j}$ and $\alpha_{k}$ are equally spaced on the log scale.
Specifically, we create the grid with $(J+2)(K + 2)$ points by first selecting maximum values $\lambda_{\textrm{max}}$ and $\alpha_{\textrm{max}}$ and minimum nonzero values $\lambda_{\textrm{min}}$ and $\alpha_{\textrm{min}}$ for $\lambda$ and $\alpha$, respectively. Then, the $\lambda_{j}$ are set to $\lambda_{0} = 0$ and $\lambda_{j} = \lambda_{\textrm{min}}\exp\big\{ (j-1)\log(\lambda_{\textrm{max}}/\lambda_{\textrm{min}})/J \big\}$, for $j \geq 1$,
and the $\alpha_{k}$ are defined as $\alpha_{0} = 0$ with $\alpha_{k} = \alpha_{\textrm{min}}\exp\big\{ (k-1)\log(\alpha_{\textrm{max}}/\alpha_{\textrm{min}})/K \big\}$, for $k \geq 1$.

\textbf{AIC-based criterion.} 
A potential drawback of using cross-validation to select $(\lambda, \alpha)$ is
that the size of the internal dataset can be quite modest, and a cross-validation selection rule for $(\lambda, \alpha)$ can have high variance and lead to poor predictive performance.
An alternative to this is to look at the effective degrees of freedom associated
with a given choice of $(\lambda, \alpha)$ and to choose these hyperparameters by minimizing an information criterion.

Traditionally, the degrees of freedom associated with a method for generating a vector of fitted values of the form $\hat{\mathbf{Y}} = \mathbf{c} + \mathbf{H}\mathbf{Y}$ is defined as $\textrm{tr}( \mathbf{H} )$. 
Motivated by this and by (\ref{eq:mm_update}), we define the 
degrees of freedom associated with any value of $(\lambda, \alpha)$ as
\begin{equation}
\textrm{df}(\lambda, \alpha) = \textrm{tr}\Bigg\{  \Big( \mathbf{X}^{T}\mathbf{X} + \alpha\mathbf{I}_{p} - \frac{\lambda}{4} \mathbf{A}_{\nu}^{T}\mathbf{W}_{0}\mathbf{A}_{\nu} \Big)^{-1}\mathbf{X}^{T}\mathbf{X} \Bigg\}. 
\label{eq:df_definition}
\end{equation}
Note that the degrees of freedom in (\ref{eq:df_definition}) equals the trace of $\mathbf{H}^{(1)}$, where $\mathbf{H}^{(1)}$ arises from 
the fitted values of the one-step estimator $\mathbf{X}\bbeta^{(1)} = \mathbf{c}^{(1)} + \mathbf{H}^{(1)}\mathbf{Y}$ where it is assumed that $\bbeta^{(0)} = \mathbf{0}$. 
Using the degrees of freedom defined in (\ref{eq:df_definition}), we define the AIC criterion associated with an estimate $\hatbbeta_{\lambda, \alpha}$ as $\textrm{AIC}(\lambda, \alpha) = 2L_{I}(\beta_{0}, \hatbbeta_{\lambda, \alpha}; \alpha) + 2\textrm{df}(\lambda, \alpha)$.

\section{Simulation Studies} \label{sec:sim_studies}
In these simulation studies, we primarily assess the ability of our ranking integration approach to improve the predictive performance of an estimated internal risk model. To this end, we evaluate the mean-squared error (MSE) $\frac{1}{n_{\textrm{test}}}\sum_{i=1}^{n_{\textrm{test}}} \{ \mu_{I}(\mathbf{x}_{i}') - \hat{\mu}_{I}(\mathbf{x}_{i}') \}^{2}$, where $\mathbf{x}_{i}', i = 1, \ldots, n_{ \textrm{test} }$ are independent draws from the same distribution as the covariates $\mathbf{x}_{i}$ the internal data, and where $\hat{\mu}_{I}(\mathbf{x})$ is an estimate of the internal risk model mean function $E_{I}\{Y_{i} \mid \mathbf{x}_{i} \}$, using the internal data $\mathcal{D}_{I} = \{(Y_{i}, \mathbf{x}_{i}); i = 1, \ldots, n_{I} \}$ with $n_{I}$ observations.

We compare our method with the following alternative methods for incorporating an externally available estimate $\bbeta_{E}$ of the regression coefficients of interest: ridge regression, a distance transfer learning (DTL) estimate, and angle-based transfer learning (ATL). The estimators of the internal risk model regression coefficients $\hatbbeta_{\textrm{ri}}(\alpha)$, $\hatbbeta_{\textrm{dtl}}(\alpha)$, $\hatbbeta_{\textrm{atl}}(\alpha, \lambda)$ for these approaches are defined as
\begin{eqnarray}
\hatbbeta_{\textrm{ri}}(\alpha) &=& (\mathbf{X}^{T}\mathbf{X} + \alpha\mathbf{I}_{p})^{-1}\mathbf{X}^{T}\mathbf{Y} \nonumber \\
\hatbbeta_{\textrm{dtl}}(\alpha) &=& (\mathbf{X}^{T}\mathbf{X} + \alpha\mathbf{I}_{p})^{-1}(\mathbf{X}^{T}\mathbf{Y} + \alpha\bbeta_{E}) \nonumber \\
\hatbbeta_{\textrm{atl}}(\alpha, \lambda) &=& (\mathbf{X}^{T}\mathbf{X} + \alpha\mathbf{I}_{p})^{-1}(\mathbf{X}^{T}\mathbf{Y} + \lambda\bbeta_{E}). \label{eq:alternative_methods}
\end{eqnarray}
In addition to assessing the performance of the three methods in (\ref{eq:alternative_methods}), we evaluate a ranking ``stacking'' approach where we add an extra column to the internal design matrix $\mathbf{X}$ containing the external rankings $r_{1}^{E}(\mathbf{x}_{1}), \ldots, r_{n}^{E}(\mathbf{x}_{n})$ and perform ordinary least squares.

\subsection{Simulation Study 1: Linear External and Internal Risk Models}

In the first simulation study, we generated an internal dataset $\mathcal{D}_{I} = \{ (Y_{i}, \mathbf{x}_{i}); i = 1, \ldots, n_{I} \}$ in each simulation replication. 
Outcomes $Y_{i}$ in $\mathcal{D}_{I}$ were generated as $Y_{i} \sim N(\mu_{I}(\mathbf{x}_{i}), \sigma_{I}^{2})$, where $\mu_{I}(\cdot)$ represents the internal data mean function. External risk scores were computed using an assumed external model mean function $\mu_{E}(\cdot)$. Both the internal and external mean functions have the form
\begin{equation}
\mu_{E}(\mathbf{z}_{i}) = \sum_{j=1}^{q} \beta_{j}^{E}z_{ij} \qquad \textrm{ and } \qquad 
\mu_{I}(\mathbf{x}_{i}) = \sum_{j=1}^{p} \beta_{j}^{I}x_{ij}, \nonumber
\end{equation}
where $\mathbf{x}_{i} = (\mathbf{z}_{i}^{T}, \mathbf{b}_{i}^{T})^{T}$.
Components of the conventional and novel covariate vectors were drawn independently as $z_{ij} \sim N(0, 1)$ and $b_{ij} \sim N(0, 1)$, but the random variables $z_{ik}$ and $b_{ij}$ are not necessarily independent.
In each simulation replication, external rankings $r_{i}^{E}(\mathbf{z}_{i})$ for the internal risk model were computed as
$r_{i}^{E}(\mathbf{z}_{i}) = \sum_{j=1}^{n} I\{ \mu_{E}(\mathbf{z}_{i}) \geq \mu_{E}(\mathbf{z}_{j}) \}$.

\noindent
\textbf{Simulation Study 1(a): Internal dataset only contains conventional covariates.}
We first simulate a setting where the internal dataset $\mathcal{D}_{I}$ only contains the conventional covariates. That is, $\mathbf{x}_{i} = \mathbf{z}_{i}$, where the components of $\mathbf{z}_{i}$ represent the same conventional covariates  used by the external risk model. The length of $\mathbf{z}_{i}$ is $5$ so that $p = q = 5$. 
We considered $11$ different choices for the external and internal mean functions $\mu_{E}(\cdot)$ and $\mu_{I}(\cdot)$. The expectation of the distance $\sum_{i=1}^{n_{I}} \{ \mu_{I}(\mathbf{z}_{i}) - \mu_{E}(\mathbf{z}_{i})\}^{2}$, and the expected Spearman's rank correlation between $\mu_{E}(\mathbf{z}_{i})$ and $\mu_{I}(\mathbf{z}_{i})$ is reported in Table \ref{tab:Sim1a} for each of the $11$ settings.

\noindent
\textbf{Simulation Study 1(b): Internal dataset contains novel covariates.}
For this study, we considered a similar design to Simulation Study 1(a), except that the dimension of the internal covariate vector $\mathbf{x}_{i}$ is now different from $\mathbf{z}_{i}$. Specifically, $\mathbf{x}_{i} = (\mathbf{z}_{i}^{T}, \mathbf{b}_{i}^{T})^{T}$, where $\mathbf{b}_{i} = (b_{i1}, b_{i2})^{T}$ contains two additional novel covariates $b_{i1}$ and $b_{i2}$. The external risk model, in this simulation study, take the form $f_{E}(\mathbf{z}_{i}) = \mathbf{z}_{i}^{T}\bbeta_{z, E}$. To induce correlation between the novel and conventional covariates, internal-dataset novel covariates were simulated as $b_{i1} = 0.4z_{i1} + e_{i1}$ and $b_{i2} = 0.25z_{i1} + 0.5z_{i3} + 0.1z_{i4} + e_{i2}$, where $e_{ij} \sim N(0, 1)$.

Because $\mathbf{x}_{i}$ contains novel covariates, we also evaluate the performance of ? with the use of marginalized ranking parameters (\ref{eq:marg_ranking_pars_sample}) in addition to the ranking parameters defined by (\ref{eq:ranking_pars}).
To sample the novel covariates $\mathbf{b}_{i}$ for creating the marginalized parameters, we use the conditional distribution $\mathbf{b}_{i}|\mathbf{z}_{i} \sim N(\bSigma_{bz}\mathbf{z}_{i}, \mathbf{I}_{2} - \bSigma_{bz}\bSigma_{bz}^{T})$, where, in each simulation replication, $\bSigma_{bz} = E( \mathbf{b}_{i}\mathbf{z}_{i}^{T} )$ is estimated from the observed values of $\mathbf{b}_{i}$ and $\mathbf{z}_{i}$. 

In this simulation study, both the DTL and ATL estimators used the target vector $\bbeta_{E}$ of regression coefficients, where the coefficients associated with the conventional covariates are set to $\bbeta_{z,E}$, while those of the novel covariates are set to $0$. Specifically, $\bbeta_{E} = (\bbeta_{z,E}^{T},\mathbf{0}^{T})^{T}$.

\noindent
\textbf{Simulation 1 Results.}
Table \ref{tab:Sim1a} presents the MSE results for each of the 11 settings of Simulation Study 1(a). In settings with substantial rank correlation between $\mu_{E}$ and $\mu_{I}$ and a small expected distance $||\mu_{E} - \mu_{I}||^{2}$ between the external and internal mean functions (i.e., Settings 1, 2, 4, and 6), distance transfer learning (DTL) performs very well relative to competing methods. In contrast, both RASPER and angle transfer learning (ATL) exhibit stronger MSE performance in settings with substantial rank correlation and a large expected distance between the mean functions (i.e., Settings 3, 5, and 7). Among these three settings, ATL had better performance when rank correlation was close to one (Setting 3), while RASPER had superior MSE performance to ATL in the two settings where rank correlation was substantial but not as close to $1$ (Settings 5 and 7). Ridge regression performed the best in settings with modest or near-zero rank correlation (i.e., Settings 8,9,10,11), but the RASPER method was very competitive with ridge regression in these settings. This suggests there is a very modest reduction in performance when using RASPER, even in settings where there is little association between the external and internal mean functions. In all settings, choosing the hyperparameters for RASPER using leave-one-out cross-validation performed superior to hyperparameter selection using AIC.
\begin{table}
\small
\centering
\begin{tabular}[h]{cccccccccc}
\toprule
\cmidrule(l{3pt}r{3pt}){3-4}\cmidrule(l{3pt}r{3pt}){5-6}\cmidrule(l{3pt}r{3pt}){7-8}\cmidrule(l{3pt}r{3pt}){9-10} 
\multirow{2}{*}{} &  &  &  &  &  &   &  &  & $\textrm{RASPER}_{S}$ \\
Setting & $\textrm{RC}(\mu_{E}, \mu_{I})$ & $|| \mu_{E} - \mu_{I} ||^{2}$ & $\textrm{RASPER}_{S}$ & $\textrm{RASPER}_{K}$ & Ridge & DTL  & ATL & Stacking  & (AIC) \\
\midrule 
  1 & 1.00 & 0.0 & 0.63 & 0.45 & 0.76 & $\mathbf{0.11}$ & 0.27 & 1.40 & 0.88 \\ 
  2 & 0.84 & 2.1 & 0.72 & 0.68 & 0.81 & $\mathbf{0.58}$ & 0.62 & 1.29 & 0.92 \\ 
  3 & 0.87 & 8.3 & 0.63 & 0.61 & 0.77 & 0.90 & $\mathbf{0.55}$ & 1.30 & 0.90 \\ 
  4 & 0.69 & 3.4 & 0.87 & 0.88 & 0.93 & 0.82 & $\mathbf{0.81}$ & 1.32 & 0.94 \\ 
  5 & 0.68 & 10.2 & $\mathbf{0.70}$ & 0.74 & 0.76 & 1.03 & 0.79 & 1.24 & 0.94 \\ 
  6 & 0.52 & 3.8 & 0.77 & 0.83 & 0.80 & $\mathbf{0.74}$ & 0.86 & 1.30 & 0.95 \\ 
  7 & 0.51 & 13.1 & $\mathbf{0.84}$ & 0.88 & 0.87 & 0.95 & 0.90 & 1.37 & 0.94 \\ 
  8 & 0.28 & 4.8 & 0.88 & 0.91 & $\mathbf{0.86}$ & 0.87 & 0.98 & 1.28 & 1.00 \\ 
  9 & 0.25 & 11.4 & 0.90 & 0.92 & $\mathbf{0.89}$ & 0.96 & 0.99 & 1.25 & 0.96 \\ 
  10 & 0.08 & 6.2 & 0.82 & 0.83 & $\mathbf{0.80}$ & 0.85 & 0.87 & 1.26 & 0.99 \\ 
  11 & 0.07 & 15.9 & 0.85 & 0.87 & $\mathbf{0.84}$ & 0.94 & 0.89 & 1.31 & 0.97 \\ 
\bottomrule
\end{tabular}
\caption{Simulation Study 1(a). Relative mean-squared error is reported for each method. 
        $\textrm{RASPER}_{S}$ (AIC) denotes the RASPER method using the Spearman association measure and hyperparameter selection according to AIC.
        The relative MSE is defined as the MSE divided by the MSE obtained by least-squares estimation.
        For each of the 11 settings, the expected rank correlation $\textrm{RankCorr}(\mu_{E}, \mu_{I})$ and expected distance 
        $|| \mu_{E} - \mu_{I} ||^{2}$ between the 
        internal and external mean functions $\mu_{I}(\cdot)$ and $\mu_{E}(\cdot)$ are reported, and the relative MSE of 
        the best performing method in each setting is in bold text.} 
\label{tab:Sim1a} 
\vspace{-8pt}
\end{table}

Table \ref{tab:Sim1b} summarizes the MSE results for Simulation Study 1(b).
As in Simulation Study 1(a), both ATL and RASPER perform particularly well relative to ridge regression and DTL
in settings with high rank correlation but a large expected discrepancy between the 
internal and external mean functions (Settings 3, 5, and 7), and in the remaining settings other than Setting 1,
RASPER was either the best or very competitive with the best performing method. 
RASPER with Kendall's $\tau$ association has the best performance in Setting 3, while
RASPER with Spearman association had the best performance in Setting 7. ATL had the best performance
in Setting 5, but RASPER with both association measures had a very close MSE to that of ATL. 
In all settings, RASPER with marginalized ranking parameters performed very similarly to RASPER
with non-marginalized ranking parameters

\begin{table}
\small
\centering
\begin{tabular}[h]{cccccccccc}
\toprule
\cmidrule(l{3pt}r{3pt}){3-4}\cmidrule(l{3pt}r{3pt}){5-6}\cmidrule(l{3pt}r{3pt}){7-8}\cmidrule(l{3pt}r{3pt}){9-10} 
Setting & $\textrm{RC}(\mu_{E}, \mu_{I})$ & $|| \mu_{E} - \mu_{I} ||^{2}$ & $\textrm{RASPER}_{S}$ & $\textrm{RASPER}_{K}$ & Ridge & DTL  & ATL & Stack & $\textrm{RASPER}_{M}$ \\
\midrule 
  1 & 0.93 & 0.7 & 0.71 & 0.57 & 0.79 & $\mathbf{0.30}$ & 0.46 & 1.17 & 0.71 \\ 
  2 & 0.87 & 3.2 & 0.61 & 0.54 & 0.69 & 0.53 & $\mathbf{0.53}$ & 1.15 & 0.61 \\ 
  3 & 0.88 & 6.7 & 0.59 & $\mathbf{0.53}$ & 0.65 & 0.93 & 0.55 & 1.11 & 0.60 \\ 
  4 & 0.72 & 4.9 & $\mathbf{0.71}$ & 0.72 & 0.78 & 0.74 & 0.78 & 1.19 & 0.72 \\ 
  5 & 0.70 & 14.3 & 0.72 & 0.73 & 0.77 & 0.92 & $\mathbf{0.70}$ & 1.17 & 0.72 \\ 
  6 & 0.51 & 5.5 & 0.73 & 0.75 & 0.75 & $\mathbf{0.67}$ & 0.73 & 1.21 & 0.73 \\ 
  7 & 0.49 & 15.9 & $\mathbf{0.71}$ & 0.73 & 0.73 & 0.82 & 0.73 & 1.17 & 0.71 \\ 
  8 & 0.27 & 6.6 & 0.82 & 0.85 & 0.81 & $\mathbf{0.80}$ & 0.88 & 1.13 & 0.81 \\ 
  9 & 0.30 & 10.4 & 0.77 & 0.79 & $\mathbf{0.77}$ & 0.93 & 0.89 & 1.15 & $\mathbf{0.77}$ \\ 
  10 & 0.05 & 8.0 & 0.69 & 0.70 & $\mathbf{0.69}$ & 0.72 & 0.74 & 1.14 & 0.69 \\ 
  11 & 0.07 & 17.6 & 0.78 & 0.80 & $\mathbf{0.76}$ & 0.92 & 0.84 & 1.16 & 0.78 \\
 \bottomrule
\end{tabular}
\caption{Simulation Study 1(b). Relative mean-squared error is reported for each method. $\textrm{RASPER}_{M}$ denotes
RASPER with Spearman association and marginalized ranking parameters.
        The relative MSE is defined as the MSE divided by the MSE obtained by least-squares estimation.
        For each of the 11 settings, the expected rank correlation $\textrm{RC}(\mu_{E}, \mu_{I})$ and expected distance 
        $|| \mu_{E} - \mu_{I} ||^{2}$ between the 
        internal and external mean functions $\mu_{I}(\cdot)$ and $\mu_{E}(\cdot)$ are reported, and the relative MSE of 
        the best performing method in each setting is in bold text.} 
\label{tab:Sim1b} 
\vspace{-8pt}
\end{table}

\subsection{Simulation Study 2: A Nonlinear External Risk Model}

In this simulation study, we use $4$ conventional covariates $\mathbf{z}_{i} = (z_{i1}, \ldots, z_{i4})^{T}$ and $2$ novel covariates 
$\mathbf{b}_{i} = (b_{i1}, b_{i2})^{T}$, and we use the same correlation structure between $\mathbf{z}_{i}$ and $\mathbf{b}_{i}$ as in Simulation Study 1(b). In all settings of this study, the mean function for the internal study is linear; specifically, $\mu_{I}(\mathbf{x}_{i}) = \sum_{j=1}^{4}\beta_{j}x_{ij}$,
where $\mathbf{x}_{i} = (\mathbf{z}_{i}^{T}, \mathbf{b}_{i}^{T})^{T}$.
In contrast, the external risk score functions $\mu_{E}(\cdot)$ are nonlinear, and in all settings considered, $\mu_{E}(\cdot)$ has the form
\begin{equation}
\mu_{E}(\mathbf{z}_{i}) = \Big(1 + f_{1}(z_{i1}) + \theta_{2}f_{2}(\mathbf{z}_{i2}) + \theta_{3}f_{1}(z_{i1})f_{2}(z_{i2}) \Big)\exp\Big\{ -\theta_{4}I(z_{i3} < 2) + 10\theta_{4}I(z_{i3} \geq 2) - \theta_{5}z_{i5}\Big\}, \nonumber
\end{equation}
where the functions $f_{j}$ are defined as $f_{1}(u) = 1/(1 + e^{-u}) - 1/(1 + e^{1 - u})$ and $f_{2}(u) = 0.5(u - 2)^{2}I(u < 7) + 12.5I(u \geq 7)$.

Because the external risk model is not a linear function of the conventional covariates, the DTL and ATL estimators cannot be directly applied in this setting. Nevertheless, we can consider modified versions of DTL and ATL where the risk score $\mathbf{Z}\bbeta_{E}$
associated with the target vector $\bbeta_{E}$ matches the true external risk 
scores as closely as possible. Specifically, we set
$\bbeta = (\tilde{\bbeta}_{z}^{T}, \mathbf{0}^{T})^{T}$, where $\tilde{\bbeta}_{z} = (\mathbf{Z}^{T}\mathbf{Z})^{-1}\mathbf{Z}^{T}\bmu_{E}$ and where $\bmu_{E} = (\mu_{E}(\mathbf{z}_{1}), \ldots, \mu_{E}(\mathbf{z}_{n}))^{T}$.

\textbf{Simulation 2 Results.}
Table \ref{tab:Sim2} summarizes the MSE performance across each of the 10 settings of Simulation Study 2. RASPER with marginalized ranking parameters had the best MSE performance in every setting with a rank correlation between the internal and external mean functions that was greater than $0.40$. In all settings, RASPER with non-marginalized ranking parameters performed very close to RASPER with marginalized parameters,
and RASPER had the second best MSE performance whenever the rank correlation exceeded $0.40$. The consistent, strong MSE performance of RASPER in Settings 1-6 demonstrates the greater flexibility of RASPER to utilize risk rankings from a nonlinear external risk model, compared to the alternative methods that shrink the fitted values directly towards a linear regression surface.
In Settings 7 and 8, where rank correlation was a more modest 0.35, RASPER still either had the best MSE performance (Setting 7) or was extremely close to the best performing method (Setting 8). In settings with low rank correlation (Settings 9 and 10), ridge regression and DTL achieved the lowest MSE, but RASPER was quite close to these methods, demonstrating that RASPER does not lead to substantial reductions in predictive performance in settings where there is little rank correlation between the internal and external risk models.

\begin{table}
\small
\centering
\begin{tabular}[h]{cccrrrrrr}
\toprule 
Setting & $\textrm{RC}(\mu_{E}, \mu_{I})$ & $|| \mu_{E} - \mu_{I} ||^{2}$ & $\textrm{RASPER}_{S}$ & Ridge & DTL  & ATL & Stack & $\textrm{RASPER}_{M}$ \\
\midrule 
  1 & 0.78 & 3.2 & 0.46 & 0.59 & 0.61 & 0.53 & 1.17 & $\mathbf{0.45}$ \\ 
  2 & 0.77 & 7.2 & 0.66 & 0.76 & 0.76 & 0.67 & 1.13 & $\mathbf{0.66}$ \\ 
  3 & 0.65 & 10.2 & 0.67 & 0.85 & 0.86 & 0.77 & 1.14 & $\mathbf{0.67}$ \\ 
  4 & 0.64 & 13.3 & 0.73 & 0.84 & 0.83 & 0.73 & 1.19 & $\mathbf{0.73}$ \\ 
  5 & 0.52 & 44.6 & 0.87 & 0.97 & 0.95 & 0.97 & 1.14 & $\mathbf{0.87}$ \\ 
  6 & 0.42 & 15.5 & 0.84 & 0.88 & 0.93 & 0.89 & 1.16 & $\mathbf{0.84}$ \\ 
  7 & 0.35 & 35.1 & 1.05 & 1.05 & 1.05 & 1.08 & 1.16 & $\mathbf{1.05}$ \\ 
  8 & 0.35 & 6.8 & 0.57 & 0.57 & $\mathbf{0.56}$ & 0.59 & 1.18 & 0.56 \\ 
  9 & 0.13 & 5.5 & 0.43 & $\mathbf{0.41}$ & 0.41 & 0.43 & 1.16 & 0.43 \\ 
  10 & 0.12 & 9.6 & 0.84 & 0.82 & $\mathbf{0.81}$ & 0.85 & 1.20 & 0.83 \\ 
 \bottomrule
\end{tabular}
\caption{Simulation Study 2. Relative mean-squared error is reported for each method. 
        The relative MSE is defined as the MSE divided by the MSE obtained by least-squares estimation.
        For each of the 10 simulation settings, the expected rank correlation $\textrm{RC}(\mu_{E}, \mu_{I})$ and expected distance 
        $|| \mu_{E} - \mu_{I} ||^{2}$ between the 
        internal and external mean functions $\mu_{I}(\cdot)$ and $\mu_{E}(\cdot)$ are reported, and the relative MSE of 
        the best performing method in each setting is in bold text.} 
\label{tab:Sim2} 
\vspace{-8pt}
\end{table}

\section{Application: A prognostic model for prostate cancer patients receiving immunotherapy} \label{sec:application}

In recent years, immune checkpoint inhibitors (ICIs) have revolutionized the treatment of many cancers. However, their use in prostate cancer has been more limited, and only a smaller fraction of patients possessing certain molecular signatures are recommended to receive treatment with an ICI \citep{lanka2023}. Because the proportion of patients with such molecular signatures is quite modest, the available sample size to construct a risk model for prostate cancer patients receiving an ICI will typically be small. Fortunately, a wide range of prognostic models for metastatic castration-resistant prostate cancer (mCRPC) have been developed and validated with larger cohorts of patients (e.g, \cite{armstrong2010}, \cite{halabi2014}), and the risk ranking information from such prognostic models has the potential to improve estimation of an ICI-specific prognostic model. 

To construct our internal risk model, we utilize the MSK-CHORD dataset \citep{jee2024}.
This is a harmonized clinicogenomic data source that records patient outcomes across a large group of patients ($n = 24,490$) and five cancer types. This data source contains 3,211 participants with prostate cancer, but only $n = 129$ of these patients ever received treatment with an ICI.
After restricting our analysis to patients who had chemotherapy prior to initiation of an ICI and
who had available tumor sequencing information, we had 79 patients with which to construct our internal
risk model.

\textbf{Conventional Covariates and External Risk Model.} To select an appropriate external risk model, we restricted our analysis to those who first received an ICIs after undergoing chemotherapy in mCRPC, which has been
the primary context in which ICIs have been administered. There are a number of widely used and well-validated risk models that apply to mCRPC prostate cancer patients after receiving first-line chemotherapy (e.g., \cite{halabi2013}) 
However, due to the limitations of the MSK-CHORD data source, we selected an external risk model that does not require as many lab measurements as that in \citep{halabi2013}. Instead, we considered simpler risk models (e.g., \cite{smaletz2002} or \cite{templeton2014}) that use fewer lab measurements. One such risk model is the more recent nomogram described in \cite{suzuki2025}, which only requires two lab measurements in addition
to common clinical covariates that are used in many other mCRPC risk models. Specifically, the following clinical covariates are needed to compute the risk score in \cite{suzuki2025}: the level of prostate specific antigen (PSA) at the start of treatment, visceral metastasis (i.e., cancer presence in either the liver, lungs, pleura, or central nervous system), Eastern Cooperative Oncology Group (ECOG) performance status score, and disease progression on prior chemotherapy. The specific values of the external risk scores $f_{E}(\mathbf{z}_{i})$ in \cite{suzuki2025} are calculated as
\begin{align}
f_{E}(\mathbf{z}_{i}) &= 0.74I(\textrm{PSA}_{i} > 30\tfrac{\textrm{ng}}{\textrm{ml}}) + 0.49I(\textrm{visceral mets}) + 0.65I(\textrm{ECOG}_{i} \geq 2) \nonumber \\
&+ 0.45\Big( 2 - \min\Big\{ 2, \frac{ \textrm{days to progression on prior chemotherapy} }{180} \Big\} \Big). \nonumber
\end{align}

\textbf{Novel Covariates.} We included several molecular features recorded in MSK CHORD that are likely to be relevant  in stratifying patient responses to immunotherapy. These include microsatellite instability (MSI), mismatch repair deficiency (MMRd), and tumor mutational burden (TMB) which are all factors that have been reported as valuable biomarkers for predicting patient response to an ICI \citep{abida2019}. 
In addition, we used an indicator of whether there was an alteration in one of the three common tumor-suppressor genes (TSG): \textit{TP53}, \textit{PTEN}, and \textit{RB1}, because these genes are known to play an important role in prostate cancer prognosis. Moreover, we include \textit{CDK12} as an additional genomic marker because it has been 
noted as an important contributor to immunotherapy response by multiple researchers (e.g., \cite{antonarakis2020cdk12}).

\textbf{Analysis and Results.} We used overall survival (OS), measured from the time of ICI treatment start, as the endpoint of interest. To enable a least-squares criterion for the local objective function, we used restricted mean survival times (RMST) ``pseudovalues'' \citep{andersen2004} as proxy outcomes. The pseudovalue for the $i^{th}$ patient is a jackknife-based quantity whose expectation equals the conditional expectation of the RMST among a population with the same covariate vector $\mathbf{x}_{i}$ as the $i^{th}$ patient. Specifically, the $i^{th}$ pseudovalue $V_{i}$ is defined as
$V_{i, \tau} = n\hat{\mu}_{\tau} - (n-1)\hat{\mu}_{\tau}^{(-i)}$, where
$\hat{\mu}_{\tau}$ is an estimate of the RMST using all observations, and
$\hat{\mu}_{\tau}^{(-i)}$ is an estimate of the RMST using all observations 
except for the $i^{th}$ observation. The RMST is defined as the expectation
of $\min\{ T_{i}, \tau\}$, where $T_{i}$ is the time to the survival endpoint of
interest, and $\tau$ is a truncation point chosen to increase the stability of the RMST estimate. In our analysis, we set the RMST truncation point to 36 months
after the start of ICI. When using the RMST pseudovalues, our local objective function is $\sum_{i=1}^{n}( V_{i,\tau} - \beta_{0} - \mathbf{x}_{i}^{T}\bbeta)^{2}$. 


Table \ref{tab:risk_model} presents the regression coefficients 
for internal RMST risk models generated from the following different 
approaches: ordinary least squares (OLS), ridge regression, 
DTL, ATL, and RASPER (using both Spearman and Kendall's $\tau$-based penalization). 
Note that the external risk model of \cite{suzuki2025} is on the log-hazard ratio scale, and direct
use of these coefficients for an RMST regression model is inappropriate. Nevertheless,
the risk rankings from this external model can provide useful information that can be harnessed by RASPER.
As is apparent from Table \ref{tab:risk_model}, estimates from ridge regression and DTL
substantially shrink the OLS coefficient estimates towards zero. 
One notable difference between OLS and ridge regression and the methods which largely use the ordering information of the external risk model (i.e., ATL and RASPER) is the role of ECOG status. The reversal of the sign of the ECOG coefficient for OLS is attributable
to the very small number of patients with $\textrm{ECOG} \geq 2$ in the MSK-CHORD cohort. In contrast to OLS and ridge regression, the well-known role of ECOG in increasing risk is preserved by ATL and $\textrm{RASPER}_{K}$, and its role is estimated to be nearly zero by $\textrm{RASPER}_{S}$.
For novel covariates where the external model provides no information, RASPER shrinks the coefficients for these covariates more strongly towards zero than it does for the conventional covariates, and hence, the coefficient estimates that RASPER generates for the novel covariates are much closer to those of ridge regression. 


\begin{table}
\small
\centering
\begin{tabular}[h]{lrrrrrrr}
\toprule
Variable & External & OLS & Ridge & DTL & ATL & $\textrm{RASPER}_{S}$ & $\textrm{RASPER}_{K}$ \\
\midrule 
 \hline
$\textrm{PSA} > 30 \textrm{ng/mL}$ & -0.74 & -175.4 & -106.5 & -106.8 & -169.2 & -117.7 & -117.1 \\
Visceral Metastasis Yes & -0.49 & -192.1 & -123.7 & -123.9 & -135.1 & -122.4 & -106.0 \\
ECOG $\geq 2$ & -0.65 & 24.2 & 11.0 & 10.5 & -122.9 & 0.9 & -7.3 \\
Prior Chemo TTP & -0.45 & -54.2 & -41.4 & -41.5 & -78.0 & -49.0 & -52.1 \\
MSI Unknown & 0 & -157.3 & -72.4 & -72.4 & -25.9 & -71.8 & -57.9 \\
MSI Unstable & 0 & -81.1 & 6.9 & 6.8 & 2.5 & 8.7 & 10.9 \\
MMRd & 0 & -106.0 & -51.5 & -51.5 & -18.6 & -51.8 & -45.2 \\
TMB & 0 & 12.9 & 11.1 & 11.1 & 10.4 & 10.8 & 10.7 \\
TSG & 0 & -89.2 & -63.2 & -63.2 & -32.3 & -59.4 & -46.8 \\
\textit{CDK12} & 0 & 142.8 & 39.4  & 39.4 & 16.9 & 34.8 & 22.0 \\
 \bottomrule
\end{tabular}
\caption{Estimated regression coefficients for 3-year RMST risk models for prostate cancer patients receiving an immune checkpoint inhibitor after first-line chemotherapy in metastatic castration-resistant prostate cancer (mCRPC) using outcomes from the MSK-CHORD data.} 
\label{tab:risk_model} 
\vspace{-8pt}
\end{table}

\begin{figure}
\centering
     \includegraphics[width=5.8in,height=3.7in]{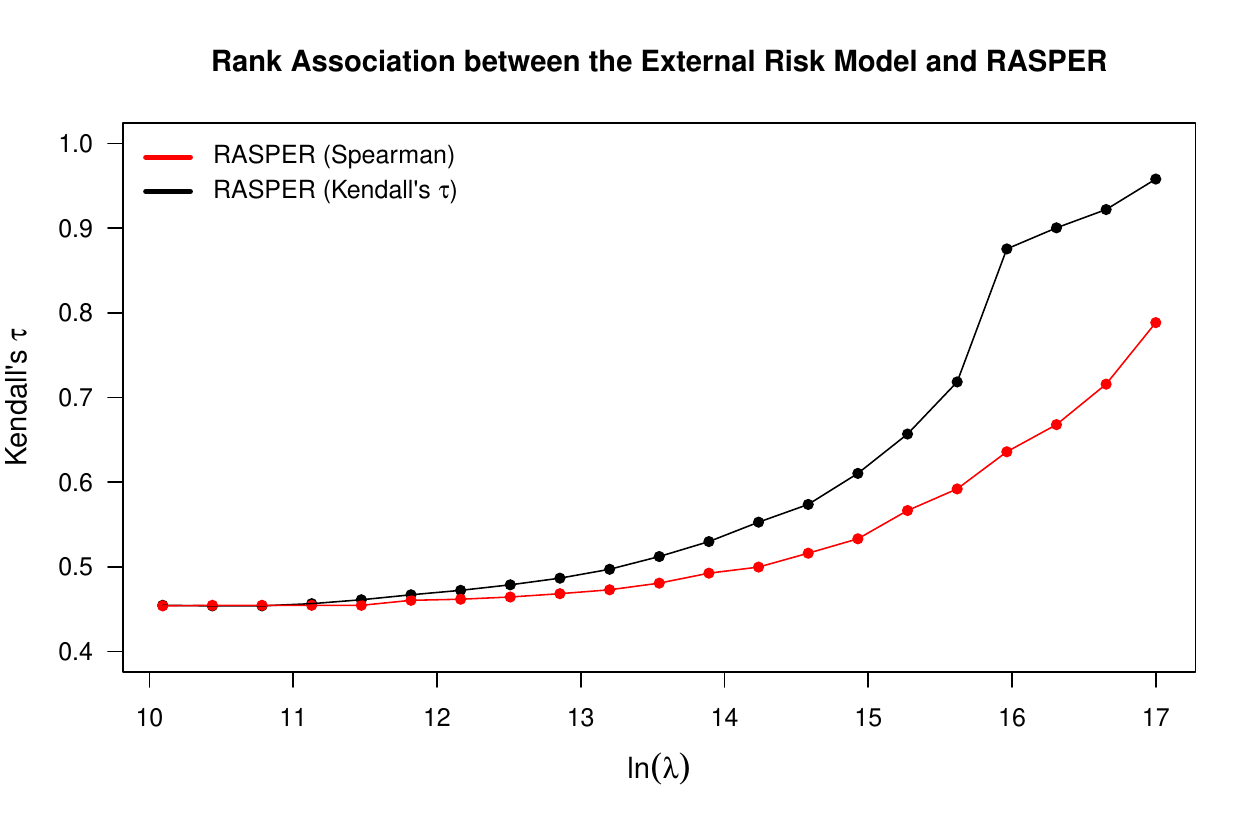}
\caption{MSK-CHORD cohort: rank association between the fitted values from RASPER and the fitted values from the external model across different choices of $\lambda$. Rank association is measured by Kendall's $\tau$. For each fixed value of $\lambda$, the hyperparameter $\alpha$ is selected using leave-one-out cross-validation.}
\label{fig:trace_plot}
\end{figure}

Figure \ref{fig:trace_plot} shows the value of the tuning parameter $\lambda$ that would lead to very strong concordance between RASPER and the external model. Specifically, this figure plots Kendall's-$\tau$ between the fitted RASPER model and the external model fitted values as the ranking penalization term $\lambda$ varies from $e^{10}$ to $e^{17}$. For each value of $\lambda$, the value of the shrinkage term $\alpha$ is selected by LOOCV. Thus, when $\lambda = 0$, RASPER corresponds to ridge regression with LOOCV selection of the tuning parameter. As the ranking penalty term approaches $0$, Kendall's $\tau$ between RASPER and the external model converges to approximately 0.45.

\begin{figure}
\centering
     \includegraphics[width=5.8in,height=3.8in]{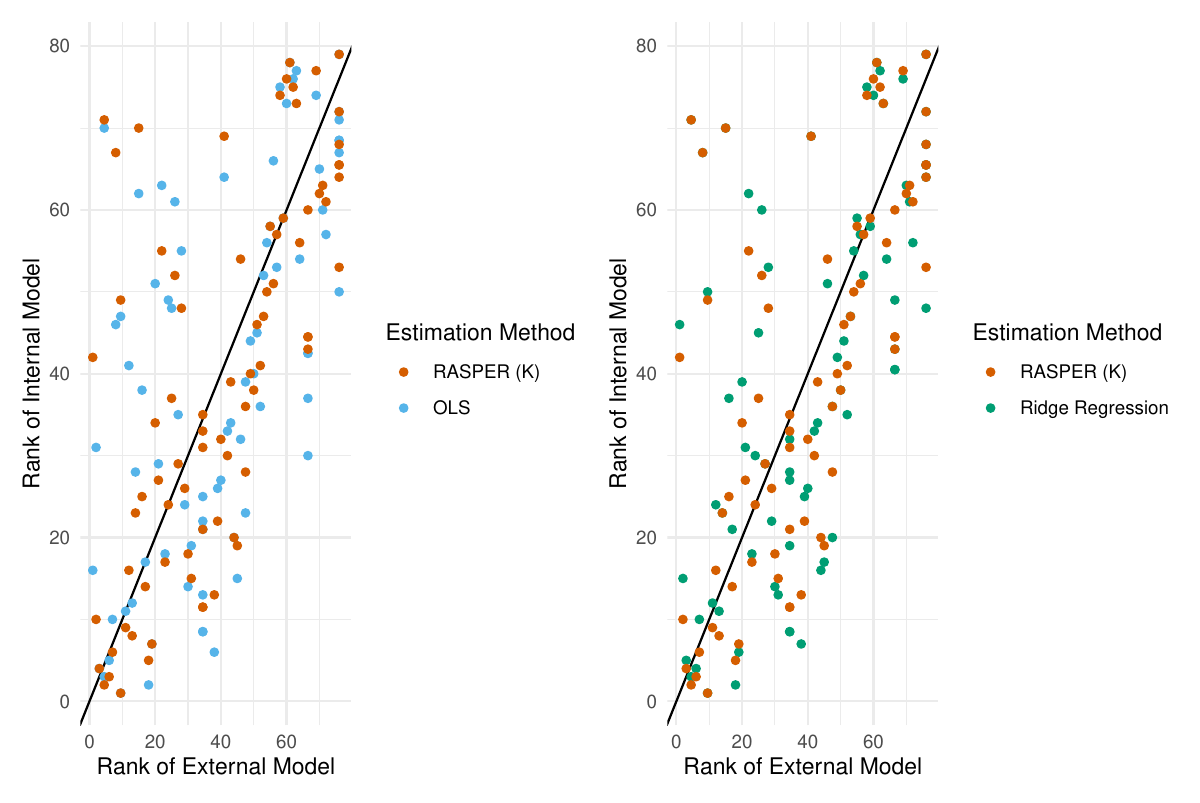}
\caption{MSK-CHORD cohort: internal model risk rankings versus external rankings for the following methods: OLS, ridge regression, and RASPER using the Kendall association appraoch.}
\label{fig:rank_agree}
\end{figure}

Figure \ref{fig:rank_agree} shows the risk rankings of OLS, ridge regression, and RASPER versus the risk rankings from applying the external model of \cite{suzuki2025}. While there is not a dramatic difference in the concordance of RASPER with the external model when compared with OLS and ridge regression, this figure shows that the risk rankings from RASPER
tend to have noticeably greater agreement with the external model risk rankings.

\section{Discussion} \label{sec:discussion}
In this article, we focused on the setting where the internal risk model uses a direct linear combination of the internal data covariates. However, because our ranking penalized approach only depends on performing comparisons with a fixed set of external risk score rankings, it can easily be used to construct more flexible internal risk models that use, for example, regression splines and generalized additive models. In such cases, the internal model covariates would simply be transformed values of the original conventional and novel covariates, and the ranking parameters (\ref{eq:rankingpar_def}) would be directly defined using this transformed set of covariates. Because the regression penalization is performed with respect to these ranking parameters, one would not need to modify the external risk score rankings or the form of the rank association measures (\ref{eq:general_disc_form}) in any way.

In our simulation studies, we have mostly observed minor differences between using marginalized ranking parameters and non-marginalized ranking parameters. This is largely due to the fact that the non-marginalized ranking parameters can be viewed as a one-sample approximation of the marginalized ranking parameters rather than an approximation based on multiple draws from the conditional distribution of the novel covariates given the conventional covariates. In practice, this means that the difference between the two forms of ranking parameters is rather modest for many candidate values of the internal regression parameters $\bbeta$, and these differences are even more muted when evaluating their impact on the estimated regression coefficient in the final estimated internal risk model.






\bibliographystyle{chicago}
\bibliography{rankintegration}

\newpage





\appendix

\section{Further Details for Rank Association Measures}
\subsection{Spearman's Rank Correlation}
If there are no ties among the ranking parameters $\psi_{1}(\bbeta), \ldots, \psi_{n}(\bbeta)$, then 
the ranking parameters $\psi_{1}(\bbeta), \ldots, \psi_{n}(\bbeta)$ exhaust the set of intergers $\{1, \ldots, n\}$ and
we have
\begin{align}
\sum_{i=1}^{n}\psi_{i}(\bbeta)\bar{r} &= \bar{r}\sum_{i=1}^{n} i = \frac{\bar{r}n(n+1)}{2} \nonumber \\ 
\sum_{i=1}^{n} \left\{ \psi_{i}(\bbeta) - \bar{\psi}(\bbeta) \right\}^{2} 
&= \sum_{i=1}^{n} i^{2} - 2\Big(\frac{n+1}{2}\Big)\sum_{i=1}^{n} i + \frac{n(n+1)^{2}}{4} \nonumber \\
&= \frac{n(n+1)(2n + 1)}{6} - \frac{n(n+1)^{2}}{2} + \frac{n(n+1)^{2}}{4} \nonumber \\
&= \frac{2n(n+1)(2n + 1)}{12} - \frac{3n(n+1)^{2}}{12}  \nonumber \\
&= \frac{4n^{3} + 6n^{2} + 2n}{12} - \frac{3n^{3} - 6n^{2} - 3n}{12}  \nonumber \\
&= \frac{n^{3} - n}{12}. \nonumber 
\end{align}

\subsection{Kendall's $\tau$}

The usual way to express Kendall's tau  between
the vectors $(\psi_{1}(\bbeta), \ldots, \psi_{n}(\bbeta))$ and $(r_{1}, \ldots, r_{n})$ is the following
\begin{eqnarray}
\kappa( \psi(\bbeta), \mathbf{r} ) &=& 
\frac{2}{n(n-1)} \sum_{i < j}  I\Big( \{ \psi_{j}(\bbeta) - \psi_{i}(\bbeta)\}\{r_{j} - r_{i} \} > 0 \Big) \nonumber \\
&=& \frac{1}{n(n-1)} \sum_{i=1}^{n}\sum_{j=1}^{n}  I\Big( \{ \psi_{i}(\bbeta) - \psi_{j}(\bbeta)\}\{r_{i} - r_{j} \} > 0 \Big) \nonumber
\end{eqnarray}
Then, if we use the fact that $\psi_{i}(\bbeta) - \psi_{j}(\bbeta) > 0$ only if $\mathbf{x}_{i}^{T}\bbeta > \mathbf{x}_{j}^{T}\bbeta$, we can see that, if there are no ties among the $\mathbf{x}_{i}^{T}\bbeta$, we have
\begin{eqnarray}
&& I\Big( \{ \psi_{i}(\bbeta) - \psi_{j}(\bbeta)\}\{r_{i} - r_{j} \} > 0 \Big) = I\Big( \{ (\mathbf{x}_{i} - \mathbf{x}_{j})^{T}\bbeta \}\{r_{i} - r_{j} \} > 0 \Big) \nonumber \\
&=& I\{ (\mathbf{x}_{i} - \mathbf{x}_{j})^{T}\bbeta > 0\}
I( r_{i} - r_{j} > 0 ) + 
I\{ (\mathbf{x}_{i} - \mathbf{x}_{j})^{T}\bbeta < 0\}
I( r_{i} - r_{j} < 0 ) \nonumber \\
&=& I\{ (\mathbf{x}_{i} - \mathbf{x}_{j})^{T}\bbeta > 0\}
I( r_{i} - r_{j} > 0 ) + 
\Big[ 1 - I\{ (\mathbf{x}_{i} - \mathbf{x}_{j})^{T}\bbeta > 0\} \Big]
I( r_{i} - r_{j} < 0 ) \nonumber \\
&=& I(r_{i} - r_{j} < 0) + 
I\{ (\mathbf{x}_{i} - \mathbf{x}_{j})^{T}\bbeta > 0\}\Big[ I( r_{i} - r_{j} > 0 ) - I( r_{i} - r_{j} < 0 ) \Big] \nonumber \\
&=& I(r_{i} - r_{j} < 0) + 
I\{ (\mathbf{x}_{i} - \mathbf{x}_{j})^{T}\bbeta > 0\}\Big[ 2I( r_{i} - r_{j} > 0 ) - 1 \Big], \nonumber 
\end{eqnarray}
where above we used the fact that if there are no ties among the $\mathbf{x}_{i}^{T}\bbeta$ we have
$I\{ (\mathbf{x}_{i} - \mathbf{x}_{j})^{T}\bbeta > 0\} = 1 - I\{ (\mathbf{x}_{i} - \mathbf{x}_{j})^{T}\bbeta < 0\}$. 

Hence, when there are no ties among the $\mathbf{x}_{i}^{T}\bbeta$, Kendall's tau can be re-expressed as
\begin{eqnarray}
&& \kappa( \psi(\bbeta), \mathbf{r} )
= \frac{1}{n(n-1)}\sum_{i=1}^{n}\sum_{j=1}^{n} I\Big( \{ \psi_{j}(\bbeta) - \psi_{i}(\bbeta)\}\{r_{j} - r_{i} \} > 0 \Big). \nonumber \\
&=& \frac{1}{n(n-1)}\sum_{i=1}^{n}\sum_{j=1}^{n} I(r_{i} - r_{j} < 0)
+ \frac{1}{n(n-1)}\sum_{i=1}^{n}\sum_{j=1}^{n} I\{ (\mathbf{x}_{i} - \mathbf{x}_{j})^{T}\bbeta > 0\}\Big[ 2I( r_{i} - r_{j} > 0 ) - 1 \Big]. \nonumber 
\end{eqnarray}

\section{Justification of MM Algorithm (Theorem 1)}

Letting $\mathbf{a}_{ij} = (\mathbf{x}_{i} - \mathbf{x}_{j})/\nu$, our aim is to minimize the following objective function:
\begin{eqnarray}
\ell(\beta_{0}, \bbeta) &=& L_{I}(\bbeta) - \lambda \log\Big( \sum_{i=1}^{n}\sum_{j=1}^{n} w_{ij}g_{1}(\mathbf{a}_{ij}^{T}\bbeta) \Big) \nonumber \\
&=& L_{I}(\bbeta) + \lambda L_{E}(\bbeta). \nonumber
\end{eqnarray}
To minimize this, let us first define the quasi-probabilities $V_{ij}(\bbeta)$ as
\begin{equation}
V_{ij}(\bbeta) = \frac{w_{ij}g_{1}(\mathbf{a}_{ij}^{T}\bbeta) }{ \sum_{i}\sum_{j} w_{ij}g_{1}(\mathbf{a}_{ij}^{T}\bbeta) }
= \frac{v_{ij}(\bbeta)}{ \sum_{i}\sum_{j} v_{ij}(\bbeta) }, 
\label{eq:quasi_probs}
\end{equation}
where $v_{ij}(\bbeta) =  w_{ij}g_{1}(\mathbf{a}_{ij}^{T}\bbeta)$.
From (\ref{eq:quasi_probs}), we can see that, for any vectors $\bbeta$ and $\bbeta^{(t)}$, we have
\begin{equation}
-\log\Big( \frac{ v_{ij}(\bbeta) }{ V_{ij}(\bbeta) }  \Big) = L_{E}(\bbeta) \qquad 
\textrm{ and } \qquad 
\sum_{i=1}^{n}\sum_{j=1}^{n} V_{ij}(\bbeta^{(t)}) = 1. \nonumber
\end{equation}
Hence, we can express $L_{E}(\bbeta)$ as
\begin{equation}
L_{E}(\bbeta) = \sum_{i=1}^{n}\sum_{j=1}^{n} V_{ij}(\bbeta^{(t)}) L_{E}(\bbeta)
= -\sum_{i=1}^{n}\sum_{j=1}^{n} V_{ij}(\bbeta^{(t)}) \log\Big( \frac{ v_{ij}(\bbeta) }{ V_{ij}(\bbeta) }  \Big).
\label{eq:external_obj_rewrite}
\end{equation}
Further expanding (\ref{eq:external_obj_rewrite}), we have
\begin{eqnarray}
L_{E}(\bbeta) 
&=& -\sum_{i=1}^{n}\sum_{j=1}^{n} V_{ij}(\bbeta^{(t)}) \log\Big( v_{ij}(\bbeta) \Big)
-\sum_{i=1}^{n}\sum_{j=1}^{n} V_{ij}(\bbeta^{(t)}) \log\Big( \frac{ 1 }{ V_{ij}(\bbeta) }  \Big) \nonumber \\
&=& -\sum_{i=1}^{n}\sum_{j=1}^{n} V_{ij}(\bbeta^{(t)}) \log\Big( v_{ij}(\bbeta) \Big)
-\sum_{i=1}^{n}\sum_{j=1}^{n} V_{ij}(\bbeta^{(t)}) \log\Big( \frac{ V_{ij}(\bbeta^{(t)}) }{ V_{ij}(\bbeta) }  \Big) \nonumber \\
&& + \sum_{i=1}^{n}\sum_{j=1}^{n} V_{ij}(\bbeta^{(t)}) \log\Big( V_{ij}(\bbeta^{(t)})  \Big) \nonumber \\
&\leq& -\sum_{i=1}^{n}\sum_{j=1}^{n} V_{ij}(\bbeta^{(t)}) \log\Big( v_{ij}(\bbeta) \Big)
+ \sum_{i=1}^{n}\sum_{j=1}^{n} V_{ij}(\bbeta^{(t)}) \log\Big( V_{ij}(\bbeta^{(t)}) \Big) \nonumber \\
&=& \sum_{i=1}^{n}\sum_{j=1}^{n} V_{ij}(\bbeta^{(t)})\log\Big( 1 + \exp\{ -\mathbf{a}_{ij}^{T}\bbeta \} \Big) - \sum_{i}\sum_{j}V_{ij}(\bbeta^{(t)})\log(w_{ij}) \nonumber \\
&& + \sum_{i=1}^{n}\sum_{j=1}^{n} V_{ij}(\bbeta^{(t)}) \log\Big( V_{ij}(\bbeta^{(t)}) \Big),
\label{eq:first_inequality}
\end{eqnarray}
where the above inequality follows from Gibb's inequality. 

To further develop the inequality for $L_{E}(\bbeta)$ in (\ref{eq:first_inequality}), we consider the following inequality for logistic log-likelihoods established in \citep{jaakkola2000}:
\begin{eqnarray}
\log\Big(1 + \exp\{ -\mathbf{a}_{ij}^{T}\bbeta \} \Big)
&\leq& \frac{1}{2}\mathbf{a}_{ij}^{T}(\bbeta^{(t)} - \bbeta)
+ \frac{1}{2}W_{ij}(\bbeta^{(t)})\Big( (\mathbf{a}_{ij}^{T}\bbeta^{(t)})^{2} - (\mathbf{a}_{ij}^{T}\bbeta)^{2} \Big) \nonumber \\
&& + \log\Big(1 + \exp\{ -\mathbf{a}_{ij}^{T}\bbeta^{(t)} \} \Big),
\label{eq:pgamma_inequality}
\end{eqnarray}
which holds for any vectors $\bbeta$ and $\bbeta^{(t)}$.
In (\ref{eq:pgamma_inequality}), the term $W_{ij}(\bbeta^{(t)})$ is defined as
\begin{equation}
W_{ij}(\bbeta^{(t)}) = \frac{\tanh( -\mathbf{a}_{ij}^{T}\bbeta^{(t)}/2)}{2\mathbf{a}_{ij}^{T}\bbeta^{(t)}}. \nonumber
\end{equation}
Combining (\ref{eq:first_inequality}) and (\ref{eq:pgamma_inequality}), we obtain
\begin{eqnarray}
L_{E}(\bbeta)
&\leq& -\frac{1}{2}\sum_{i}\sum_{j}V_{ij}(\bbeta^{(t)})\mathbf{a}_{ij}^{T}\bbeta 
- \frac{1}{2}\sum_{i=1}^{n}\sum_{j=1}^{n} V_{ij}(\bbeta^{(t)})W_{ij}(\bbeta^{(t)})(\mathbf{a}_{ij}^{T}\bbeta)^{2}
+ \frac{1}{2}\sum_{i=1}^{n}\sum_{j=1}^{n}V_{ij}(\bbeta^{(t)})\mathbf{a}_{ij}^{T}\bbeta^{(t)} \nonumber \\
&& + \frac{1}{2}\sum_{i=1}^{n}\sum_{j=1}^{n} V_{ij}(\bbeta^{(t)})W_{ij}(\bbeta^{(t)})(\mathbf{a}_{ij}^{T}\bbeta^{(t)})^{2}
+ \sum_{i=1}^{n}\sum_{j=1}^{n} V_{ij}(\bbeta^{(t)})\log\Big(1 + \exp\{ -\mathbf{a}_{ij}^{T}\bbeta^{(t)} \} \Big) \nonumber \\
&& - \sum_{i=1}^{n}\sum_{j=1}^{n}V_{ij}(\bbeta^{(t)})\log(w_{ij}) + \sum_{i=1}^{n}\sum_{j=1}^{n} V_{ij}(\bbeta^{(t)}) \log\Big( V_{ij}(\bbeta^{(t)}) \Big) \nonumber \\
&=& G(\bbeta|\bbeta^{(t)}). \nonumber
\end{eqnarray}

So, we have established that $L_{E}(\bbeta) \leq G(\bbeta|\bbeta^{(t)})$ for any vectors $\bbeta$ and $\bbeta^{(t)}$.
Now, consider $G(\bbeta^{(t)}|\bbeta^{(t)})$:
\begin{eqnarray}
G(\bbeta^{(t)}|\bbeta^{(t)}) &=& \sum_{i=1}^{n}\sum_{j=1}^{n} V_{ij}(\bbeta^{(t)})\log\Big(1 + \exp\{ -\mathbf{a}_{ij}^{T}\bbeta^{(t)} \} \Big) 
- \sum_{i=1}^{n}\sum_{j=1}^{n} V_{ij}(\bbeta^{(t)})\log(w_{ij}) \nonumber \\
&& + \sum_{i=1}^{n}\sum_{j=1}^{n} V_{ij}(\bbeta^{(t)}) \log\Big( V_{ij}(\bbeta^{(t)}) \Big) \nonumber \\
&=& -\sum_{i}\sum_{j} V_{ij}(\bbeta^{(t)})\log\Big(\frac{w_{ij}}{ 1 + \exp\{ -\mathbf{a}_{ij}^{T}\bbeta^{(t)} \} } \Big) 
 + \sum_{i}\sum_{j} V_{ij}(\bbeta^{(t)}) \log\Big( V_{ij}(\bbeta^{(t)}) \Big) \nonumber \\
&=& -\sum_{i}\sum_{j} V_{ij}(\bbeta^{(t)})\log\Big( v_{ij}(\bbeta^{(t)}) \Big) + \sum_{i}\sum_{j} V_{ij}(\bbeta^{(t)}) \log\Big( V_{ij}(\bbeta^{(t)}) \Big) \nonumber \\
&=& -\sum_{i}\sum_{j} V_{ij}(\bbeta^{(t)})\log\Big( \frac{ v_{ij}(\bbeta^{(t)}) }{ V_{ij}(\bbeta^{(t)}}) \Big) \nonumber \\
&=& \sum_{i}\sum_{j} V_{ij}(\bbeta^{(t)}) L_{E}(\bbeta^{(t)}) \nonumber \\
&=& L_{E}(\bbeta^{(t)}). \nonumber 
\end{eqnarray}
We can express $G(\bbeta|\bbeta^{(t)})$ as
\begin{equation}
G(\bbeta|\bbeta^{(t)}) = -\frac{1}{2}(\mathbf{V}_{t}\mathbf{A}\bbeta)^{T}\mathbf{1} - \frac{1}{2}\bbeta^{T}\mathbf{A}^{T}\mathbf{V}_{t}\mathbf{W}_{t}\mathbf{A}\bbeta + C(\bbeta^{(t)}), \nonumber 
\end{equation}
where $\mathbf{W}_{t}$ is a matrix with diagonal elements $W_{ij}(\bbeta^{(t)})$ and
$\mathbf{V}_{t}$ is a matrix with diagonal elements $V_{ij}(\bbeta^{(t)})$.

Note that for continuous outcomes,
\begin{eqnarray}
\ell(\bbeta) &=& -\bbeta^{T}\mathbf{X}^{T}\mathbf{y} 
+ \frac{1}{2}\bbeta^{T}\mathbf{X}^{T}\mathbf{X}\bbeta - 
\frac{\lambda}{2}\bbeta^{T}\mathbf{A}^{T}\mathbf{V}_{t}\mathbf{1} + \frac{\lambda}{2}\bbeta^{T}\mathbf{A}^{T}\mathbf{V}_{t}\mathbf{W}_{t}\mathbf{A}\bbeta + \lambda C(\bbeta^{(t)}) \nonumber \\
&=& -\bbeta^{T}(\mathbf{X}^{T}\mathbf{y} + \frac{\lambda}{2}\mathbf{A}^{T}\mathbf{V}_{t}\mathbf{1})
+ \frac{1}{2}\bbeta^{T}\Big( \mathbf{X}^{T}\mathbf{X} + \lambda \mathbf{A}^{T}\mathbf{V}_{t}\mathbf{W}_{t}\mathbf{A}\Big)\bbeta + C(\bbeta^{(t)}). \nonumber
\end{eqnarray}
Thus, the MM update of the parameter vector is 
\begin{equation}
\bbeta^{(t+1)} = \Big( \mathbf{X}^{T}\mathbf{X} + \lambda\mathbf{A}^{T}\mathbf{V}_{t}\mathbf{W}_{t}\mathbf{A}\Big)^{-1}(\mathbf{X}^{T}\mathbf{y} + \frac{\lambda}{2}\mathbf{A}^{T}\mathbf{V}_{t}\mathbf{1}). \nonumber 
\end{equation}

\end{document}